\documentclass[twocolumn,
showpacs,
showkeys,
preprintnumbers,
nofootinbib,
superscriptaddress,
amsmath,
amssymb,
floatfix,
secnumarabic,
aps,
pra,
a4paper,
notitlepage,
final,
]{revtex4}%

\usepackage[colorlinks=true,urlcolor=blue,breaklinks=true]{hyperref}
\usepackage{graphicx}
\usepackage{epsfig}

\newcommand{\beq}{\begin{equation}}
\newcommand{\eeq}{\end{equation}}
\newcommand{\beqar}{\begin{eqnarray}}
\newcommand{\eeqar}{\end{eqnarray}}
\newcommand{\bal}{\begin{aligned}}
\newcommand{\eal}{\end{aligned}}

\def\dalam{\hbox
{\vrule\vbox{\hrule\hbox to 1ex{ \hfill}\kern 1 ex\hrule}\vrule}}

\def\1/2{\hbox{$ {1 \over 2}$ }}

\def\h{\hbar}
\def\i/h{{i \over \h}}

\def\inf{\infty}
 
\def\v{\vec}

\def\b{\beta} 
 
\def\g{\gamma} \def\G{\Gamma} 
 \def\D{\Delta}
\def\l{\lambda} 
 
 \def\tE{\tilde {E}}

\def\ve{\varepsilon}
\def\s{\sigma}
\def\r{\rho} 
\def\x{\xi}
\def\c{\chi} 

 \def\F{\Phi}
 
\def\p{\psi}

\def\<{\langle}
\def\>{\rangle}

\def\({\left(}
\def\[{\left[}
\def\){\right)}
\def\]{\right]}

\usepackage{subfigure}

\usepackage[notcite,notref,color]{showkeys}  

\usepackage{tikz}
\usetikzlibrary{decorations.pathreplacing}
\usetikzlibrary{decorations.pathmorphing}    
\usepackage{multirow}
\usepackage{dcolumn}		
\newcolumntype{.}{D{.}{.}{-1}}
\newcolumntype{i}[1]{D{.}{.}{#1}}

\newcommand{\myfrac}[2]{{\ifmmode{}^{#1}\!/_{\!#2}\else${}^{#1}\!/_{\!#2}$\fi}}

\begin{document}
\sloppy

\title{Atomic H  over plane: effective potential and level reconstruction}

\author{S.~Artyukova}
\email{s.artyukova@physics.msu.ru} \affiliation{Department of Physics and
Institute of Theoretical Problems of MicroWorld, Moscow State
University, 119991, Leninsky Gory, Moscow, Russia}

\author{K.~Sveshnikov}
\email{costa@bog.msu.ru} \affiliation{Department of Physics and
Institute of Theoretical Problems of MicroWorld, Moscow State
University, 119991, Leninsky Gory, Moscow, Russia}

\author{A.~Tolokonnikov}
\email{tolokonnikov@physics.msu.ru} \affiliation{Department of Physics and
Institute of Theoretical Problems of MicroWorld, Moscow State
University, 119991, Leninsky Gory, Moscow, Russia}

\date{\today}


\begin{abstract}
The behavior of atomic H in a semi-bounded space $z \geq 0$ with the condition of ``not going through'' the boundary (the surface $z=0$) for the electronic wavefunction (WF) is considered. It is shown that in a wide range  of  ``not going through'' condition parameters the effective atomic potential, treated as a function of the distance $h$ from H to the boundary plane,  reveals  a well pronounced minimum  at certain finite but non-zero $h$, which describes the mode of ``soaring'' of the atom above the plane. In particular cases of Dirichlet and Neumann conditions the analysis of the soaring effect is based on the exact analytical solutions of the problem in terms of generalized spheroidal Coulomb functions. For $h$ varying between the regions $h \gg a_B$ and $h \ll a_B$ both the deformation of the electronic WF and the atomic state are studied in detail. In particular, for the Dirichlet condition the lowest $1s$ atomic state transforms into $2p$-level with quantum numbers $210$, the first excited ones $2s$   --- into $3p$ with numbers $310$, $2p$ with $m=0$ --- into  $4f$ with numbers $430$, etc. At the same time, for  Neumann condition the whole picture of the levels transmutation changes drastically. For a more general case of Robin (third type) condition the variational estimates, based on  special type trial functions, as well as the direct numerical tools, realized by  pertinent modification of the finite element method, are used. By means of the latter it is also shown  that in the case of a sufficiently large positive affinity  of the atom to the boundary plane a significant reconstruction of the lowest levels  takes place, including the  change of both the asymptotics and the general dependence  on $h$.
   \end{abstract}


\maketitle

\section{Introduction}\label{sec:intro}

Considerable amount of theoretical and experimental activity has
been focused recently on spatially confined atoms and molecules
\cite{Jaskolski1996,Sabin2009,Sen2014}. The interest is largely due to the nontrivial physical and chemical properties that arise for quantum systems in such a state of complete or incomplete confinement. The interaction of confined particles with the environment, forming the cavity or volume boundary, is usually simulated  by means of a suitable boundary condition, imposed on their wavefunctions (WF). The pioneering works on  quantum system in a closed cavity are the Wigner-Seitz model \cite{Wigner1933,Wigner1934} and  the papers \cite{Sommerfeld1938} and \cite{Michels1937} on  atomic H in a spherical cavity. More concretely, in the Wigner-Seitz model the metallic bond formation  in alkali metals has been considered in terms of  the Neumann condition for the valence electron  on the boundary of the corresponding Wigner-Seitz cell. In Ref.\cite{Sommerfeld1938} the exact solution for  atomic H in a spherical cavity with the Dirichlet boundary condition was found, while in Ref.\cite{Michels1937} such a model has been used for description of atomic H under high pressure.

Atoms in the Euclidean half-space $\Re^3/2$ have been first explored in Ref.\cite{Levine1965}, devoted to the properties of  the impurity donor atom placed on the plane boundary of the dielectric crystal. Due to a large positive affinity  it is energetically favorable for the valence electron to reside inside the crystal that allows to  simulate the crystal boundary by means of the Dirichlet condition imposed on the electronic WF. Afterwards, this model has been actively used by solving the spectral problem for an atom, placed  inside a semiconductor near its surface. In contrast to the one considered in Ref.\cite{Levine1965}, in the latter case the analytic solution of the Schroedinger eq. (SE) is absent, therefore  various numerical tools are used (see, e.g., Refs.\cite{Liu1983,Kovalenko1992a,Shan1985,Cruz2008}).

Further research was not limited to the study of  spherical cavities or plane boundaries, rather it was motivated by the circumstance that SE with Coulomb potential allows for  separation of variables not only in spherical, but also in spheroidal and parabolic frames. Therefore besides a sphere \cite{Wigner1933,Wigner1934,Sommerfeld1938,Michels1937} and plane \cite{Levine1965,Liu1983,Kovalenko1992a,Shan1985,Cruz2008} there have been considered other ``natural'' for the above-mentioned frames surfaces, namely, elliptic cones, plane angles, etc. (see Refs.\cite{Mendez2011,LeyKoo1991,Cruz1995,LeyKoo1997,LeyKoo1993a, LeyKoo1993b,ChaosCador2005} and refs. therein), with impenetrable \cite{Liu1983,Kovalenko1992a,Shan1985,Cruz2008,Mendez2011,LeyKoo1991,Cruz1995,LeyKoo1997,LeyKoo1993a, LeyKoo1993b,ChaosCador2005} or semipermeable \cite{Shan1990,Kovalenko1992b} potential wall  used as a boundary.   It should be mentioned, however, that for such problems the explicit analytic solution in a closed form is actually impossible, since the answers are formulated in terms of convergent infinite series, and so practically meaningful calculations are restricted to a finite number of first terms in these series \cite{Liu1983,Kovalenko1992a,Shan1985,Cruz2008,Mendez2011,LeyKoo1991,Cruz1995,LeyKoo1997,LeyKoo1993a, LeyKoo1993b,ChaosCador2005,Shan1990,Kovalenko1992b}.

The behavior of atomic He and of He$^+$ ion in a half-space with plane boundary  \cite{Cruz2008} has been studied experimentally \cite{Wethekam2008,Monreal2013}. However, the performed experiments have shown that the levels shift in He depends not only on the distance to the boundary (that could be described within the idealized model with the Dirichlet boundary condition), but also on the crystal structure of matter (Al and noble metals) that forms this boundary. The Dirichlet condition cannot in principle take account for such effects, since it nullifies the electronic WF  on the boundary. At the same time, the general boundary conditions of ``not going through'' (i.e., Robin, or third kind) allow for a sufficiently more wide problem statement, which doesn't imply the  vanishing WF on the volume boundary \cite{Pupyshev2000,Sen2009,AlHashimi2012a,AlHashimi2012b,Sveshnikov2013a,Sveshnikov2013b,Sveshnikov2013c, Sveshnikov2013d,Pupyshev2014}. Moreover, such conditions are able to  take  into account the interaction of confined particles with medium, surrounding the cavity or demarcating the half-space \cite{Sveshnikov2013a,Sveshnikov2013b,Sveshnikov2013c,Sveshnikov2013d}, \cite{Sveshnikov2017a, Sveshnikov2017b}. It would be worth to note that the  term ``not going through'', used here, underlines that these conditions  do not necessarily originate from the actual confinement of particles inside the given volume, rather they may be caused by  a significantly wider number of reasons, as it takes place, in particular, in the Wigner-Seitz model of an alkali metal \cite{Wigner1933,Wigner1934}, where the valence electron state is principally delocalized.  The latter circumstance turns out to be quite important,
since in some cases the cavities, where a
particle or an atom could reside, form a lattice, similar to that of an alkali metal,  like
certain interstitial sites of a metal supercell, e.g. next-to-nearest octahedral
positions of palladium fcc lattice \cite{Caputo2003}. In this case a particle  (or a valence electron, provided that the whole lattice of cavities is occupied by identical atoms) finds itself in a periodic potential of a cubic lattice, and so the description of its ground state could be based on the first
principles of the Wigner-Seitz model  \cite{Sveshnikov2013c,Sveshnikov2013d}, \cite{Sveshnikov2017a, Sveshnikov2017b}.

In the present paper  the behavior of  atomic H in the half-space  $z \geq 0$ with the plane boundary  $z=0$, whereon the electronic WF should be subject of the general Robin condition, is studied in the adiabatic approximation with respect to nucleus motion. The main motivation is that if H is inside a spherical cavity with the radius $R$ and a  $\delta$-like potential on its border, which simulates the interaction of  atomic electron with medium, wherein the cavity is formed, then for a wide range of  surface interaction coupling constant  (including repulsion!) with growing  $R$ the equilibrium position of the atom in the center of cavity ceases to be stable and it shifts to the  border \cite{Sveshnikov2017a, Sveshnikov2017b}. When the   curvature  of the border becomes much larger than $a_B$, then  the problem of an H over plane with  Robin boundary condition for the electronic WF appears in a natural way. In this problem there are two most important questions. The first one is under which conditions imposed on the surface interaction the effective atomic potential treated as a function of the distance $h$ to the plane  reveals a minimum for finite and nonzero $h$. A preliminary, rather rough estimate for this effect has been considered in Refs.\cite{Sveshnikov2017a, Sveshnikov2017b}. The second concerns the additional ``power-like'' levels, which appear always under Robin boundary condition with attractive surface interaction and become the lowest ones, when the attraction is sufficiently strong.  Here we'll present a detailed study of both questions.

The paper is organized as follows. In Section \ref{sect:1} the general problem statement for atomic H over plane is presented, supplied with the required information  for further analysis from the similar problem in a spherical cavity. In Section \ref{sect:2}  the partial cases of the general ``not going through'' condition, namely, the Dirichlet and Neumann ones, which allow for a ``quasi-exact'' analysis of the problem, are considered. There are explored in detail the ground state and two first excited levels ``$2s$'' and ``$2p$''.
In Section  \ref{sect:3}  special kind trial functions, which allow for qualitative reproduction of the analytic results for the Neumann condition, as well as of the results of direct numerical calculations for the Robin case, with additional restriction on the magnitude of attractive surface interaction from above, are considered.
In Section \ref{sect:4} a sufficiently more wide range of the magnitude of surface interaction is explored. The numerical results for the lowest levels  are presented and  compared, if possible, with those  achieved via  variational estimates. Special attention is paid to  additional ``power-like'' levels, which come into play for sufficiently strong attractive surface interaction.
In Conclusion (Section \ref{sect:5}) the main consequences of the atomic H ``soaring'' effect and lowest levels reconstruction in the case of sufficiently strong positive  electron-plane affinity  are discussed.

Throughout the paper the a. u. $\hbar=m=e=1$, Computer Algebra Systems (such as Maple 18) to facilitate  the analytic calculations  and GNU Octave code for boosting the numerical work, are used.

\section{The problem statement for atomic  H in $\Re^3/2$}\label{sect:1}

The initial formulation of the problem repeats almost completely  the one, considered in Refs.\cite{Pupyshev2000,Sen2009,AlHashimi2012a,AlHashimi2012b,Sveshnikov2013a,Sveshnikov2013b,Sveshnikov2013c, Sveshnikov2013d,Pupyshev2014}. In the present case the energy functional for the  electronic WF  takes the form
\begin{equation}\label{eq:1_01}
E[\psi]=\int\limits_{z \geq 0} d\vec r \left[\frac{1}{2}|\vec{\nabla}\psi|^2+V(\vec r)|\psi|^2\right]+\frac{1}{2}\int\limits_{z=0} d\vec \rho\, \lambda(\vec \rho)|\psi|^2 \ ,
\end{equation}
where
\begin{equation}\label{eq:1_011}
V(\vec r)=-1/\sqrt{\rho^2 + (z-h)^2}
\end{equation}
is the Coulomb potential of the nucleus (the proton) with coordinates $(0, 0, h)$, while the surface term describes the interaction of the atomic electron on the border $z=0$, demarcating the half-space  $z > 0$, with medium, filling another half-space with $z < 0$. The concrete properties of this surface interaction are determined via a real-valued function $\lambda (\vec \rho)$.

Proceeding further, from the variational principle one obtains
\begin{equation}\label{eq:1_02}
\left[ -\frac{1}{2}\triangle + V(\vec r )  \right] \psi=E \psi
\end{equation}
for  $z>0$ and the boundary condition on the surface  $z=0$
\begin{equation}\label{eq:1_03}
\left[ \partial/\partial z - \lambda(\vec \rho )  \right] \psi \Big|_{z=0}=0 \ .
\end{equation}
The  ``not going though''  property is fulfilled
here via vanishing normal to the surface component of the  quantum-mechanical flux $\vec j$
\begin{equation}\label{eq:1_05}
\left. j_z \right|_{z=0}=0 \ .
\end{equation}
At the same time, the tangential components of  $\vec j$ could be
remarkably different from zero on the surface $z=0$ and so the atomic electron could
be found quite close to the boundary with a marked probability.

When $\l=0$, the interaction of the atom with environment is
absent and so eq. (\ref{eq:1_03}) transforms into  Neumann (second kind)
condition
\begin{equation} \label{eq:1_06}
\left.  \partial\psi/\partial z  \right|_{z=0}=0 \ .
\end{equation}
If $\l \to \infty$, then  (\ref{eq:1_03}) transforms into the Dirichlet
condition
\begin{equation} \label{eq:1_07}
\left.  \psi \right|_{z=0}=0 \ ,
\end{equation}
and hence,
describes  confinement  by an impenetrable barrier.

Here it should be noted  that the interaction of atomic electron with environment by means of the surface $\delta$-potential is a simplification of the actual situation, since in this picture  the atomic structure of medium is disregarded. So the  different type surface excitations and inside the bulk don't be  taken into account explicitly, rather they are involved into the properties of the boundary condition. However, to the first approximation such description turns out to be satisfactory, because it allows for qualitative accounting for such general features as the atomic H affinity to medium,  surrounding the cavity or demarcating the half-space. The general literature on the interaction of atomic H with surfaces of different nature is
extremely vast, since it includes such  interdisciplinary aspects as hydrogenation, dehydrogenation, and hydrogenolysis processes
that are crucially important to the chemical industry. A thorough
review of this literature is beyond the scope of the present
paper (for a detailed summary of older hydrogen literature,
the reader is referred to excellent reviews presented in Refs.\cite{Christmann1988,Zuettel2003, Fukai2005}). More recent studies on this subject are considered in Refs.(\cite{Christmann2009,Wilde2014,Callini2015,Stolten2016} and citations therein).
The nowadays most efficient self-consistent methods for the study of various properties of both surface
and subsurface atomic hydrogen  are based mostly on  Gradient corrected periodic density functional theory (DFT-GGA) slab calculations \cite{Hammer1996, Pedersen1999, Pallassana1999,Pallassana2000,Greeley2005}. Very roughly, the averaged affinity estimates (ignoring both the difference between physisorption and chemisorption and the dependence on the current concentration of H in the bulk), found this way for atomic H in terms of $\l$, vary from very large positive values (liquid He) to large negative ones in PdH$_x$ and TiH$_x$, provided  $x \ll 1$.

A simple example for estimating $\l$  follows directly from the well-known quantum-mechanical problem of a particle in the $\delta$-potential $V(\vec r)=\(\lambda/2\)\,\delta(r-R)$ \cite{Tosatti1994,Amusia1998}. In this case the ground state WF $u_0(r)=r R_0(r)$ up to normalization coefficient takes the form
\begin{equation}\label{eq:1_09}
\begin{gathered}
u_0(r)=\sqrt{\dfrac{r}{R}}K_{1/2}(\beta R)I_{1/2}(\beta r) = \\ = \frac{\exp(-\beta R)}{\beta R}\sinh(\beta r) , \quad r\leq R \ , \\
u_0(r)=\sqrt{\dfrac{r}{R}}I_{1/2}(\beta R)K_{1/2}(\beta r) = \\ = \frac{\sinh(\beta R)}{\beta R}\exp(-\beta r) , \quad r > R \ ,
\end{gathered}
\end{equation}
with $I_{1/2}(z)$ and $K_{1/2}(z)$ being the modified Bessel functions,   $\beta=\sqrt{2I}$, while $I=-E$ is the electron affinity energy to the source of the $\delta$-potential.
The relation between  $\lambda$ and the  parameters $I$ and $R$, which are determined experimentally, is obtained from the jump in the logarithmic derivative at $r=R$
\begin{equation}\label{eq:1_10}
\beta (1+\coth(\beta R))=-\lambda \; .
\end{equation}
In particular, for the negatively charged fullerene ion   $C_{60}^{-}$ one has $R=6.639\, a_B$ and  $I=2.65\, eV=0.097\, Ha$ \cite{Tosatti1994,Amusia1998}, whence $\l=- 0.885\, Ha \times a_B$.
The additional electron in $C_{60}^{-}$ the most part of time resides in those spatial regions, where  its interaction with $C_{60}$ is negligibly  small, that allows to find the electronic WF in almost the whole space without detailed information on the true  $C_{60}$ potential. Translating this result to our problem, we get the estimate for $\l$ in (\ref{eq:1_01}) as $\pm O(1)$ in units $Ha \times a_B$, which is consistent with results for a large set of transition metals and
near-surface alloys \cite{Greeley2005}.

From the beginning the problem  for the energy levels of atomic H in $\Re^3/2$ with the boundary condition (\ref{eq:1_03}) requires for taking into account the following circumstance. Because this problem partially  (but not completely!) could be considered through the limit $R \to \inf$ of the similar problem for atomic H, confined to a spherical cavity of radius  $R$ with the same boundary condition, the energy levels of H in a cavity should possess their analogues in  $\Re^3/2$. The complete correspondence here is absent, since in a spherical cavity with the nucleus placed in the center for any $R$ the orbital moment is conserved and so the eigenstates of H are labeled by the quantum numbers $lm$, although the additional degeneration of levels disappears \cite{AlHashimi2012a,AlHashimi2012b}. In $\Re^3/2$ for any finite distance $h$ between the nucleus and the boundary plane $z=0$  there remains only the axial symmetry, hence, the only conserved quantity is  $l_z$. In a cavity with finite $R$ such situation also takes place when the equilibrium position of H shifts from the center to the border, that happens whenever $\l < q$ with $q$ being the nucleus charge \cite{Sveshnikov2017a, Sveshnikov2017b}.  In $\Re^3/2$ the spherical symmetry restores only for infinite distances between the atom and  plane and only in the case, when the atomic electron is localized in the nucleus vicinity, where it falls into the eigenstates of the free atom. But this is not the general case. Namely, for $\l <0$  atomic H in a spherical cavity with finite $R$ acquires a set of qualitatively different levels, when the electron is partially (but not completely) localized   in the vicinity of the border \cite{Sveshnikov2013c,Sveshnikov2013d, Sveshnikov2017a, Sveshnikov2017b}. These states are orthogonal to the ``normal'' atomic states, in which the electron is localized in the nucleus vicinity, and  together they form the complete set of states of atomic H, trapped into a cavity, for $\l <0$. Moreover, these additional states reveal a number of principally different features, which show up most clearly in their asymptotic behavior for large separation between nucleus and the border, which turns  out to be a power-like one with the common limiting point $(-\l^2/2)$. These levels possess their own analogues in $\Re^3/2$, when the atomic electron is partially (but not completely) localized in the neighborhood of the boundary plane, and  are  qualitatively different from the ``normal'' levels, corresponding to the electron localization in the nucleus vicinity. Most clearly this difference shows up in the dependence of these levels on $h$  for   $h \to \inf$.

Since a large number  of problems concerning the one-electron atom with the nucleus charge  $q$, trapped into a spherical cavity of radius $R$ with the Robin  condition, has been already considered in Refs.\cite{Pupyshev2000,Sen2009,AlHashimi2012a,AlHashimi2012b,Sveshnikov2013a,Sveshnikov2013b,Sveshnikov2013c, Sveshnikov2013d,Pupyshev2014}, here only a certain detail clarification is required. Quite similar to the previous example with  $C_{60}$, we'll imply here that the surface interaction is determined by
a constant $\l$, while the motionless point-like atomic nucleus is placed in
the center of  cavity, hence, the  spherical symmetry is maintained. From the solution of the
 Schroedinger-Coulomb problem   for the  radial electronic WF  with the orbital momentum $l$  one obtains up  to a
numerical factor
\beq \label{f17} R_l (r) =e^{-\g r} r^l \ \F (b_l , c_l , 2 \g r) \ , \eeq
where
\beq \label{f18} \g=\sqrt{-2E} \ ,
\ b_l=l+1-q/\g \ , \ c_l=2l+2 \ , \eeq with  $\F(b,c,z)$ being the
confluent hypergeometric function of the first kind (Kummer
function). Definition, notations and main properties of the Kummer
function follow Ref.\cite{bateman1953}. The energy levels with the orbital momentum  $l$ are determined from the equation
\beq \label{f19}
\[ q/\g + (\l-\g)R -1 \]  \F_R + \[l+1 -q/\g\] \F_R(b+) =0 \ ,
\eeq
where
\beq \label{f20} \F_R=\F (b_l , c_l , 2 \g R) \ , \
\F_R(b+)=\F (b_l+1 , c_l , 2 \g R) \ . \eeq

The reconstruction of the electronic spectrum  in the cavity shows up most clearly for $R \to \inf$, when by means of the asymptotics of $\F_R \ , \ \F_R(b+)$ one obtains from (\ref{f19}), that besides the ``normal'' discrete spectrum of the free atom in the case of attractive surface interaction with $\l <0$ there emerges another set of levels $\tE_l (R)$ with a power-like asymptotics for  $R \to \inf$ and a common limiting point  $\tE_l(\inf) = - \l^2/2$, namely
\begin{multline}\label{f27}
\tE_l (R) \to - \l^2/2 +   (\l-q)/R \ + \\ +  \ \(l(l+1) -1 + q/\l\) /  R^2 + O\(1/R^3\) \ , \ R \to \inf \ .
\end{multline}
It should be mentioned that for $\l < -q < 0 $ these levels become the lowest ones for any $R$ and as functions of  $R$ behave similar to the lowest level of a particle confined to a spherical cavity \cite{Sveshnikov2013c}, namely, $\tE_l (R)$  have the form of hyperbolas shifted down relative to the x-axis. The lowest one in this bundle of power levels with different $l$ will be quite naturally the $s$-level with $l=0$, while all the others with  $l \not =0$ will be  raised higher in proportion to their centrifugal energy.

At the same time, the  ``normal''  atomic states tend  for $R \to \inf$ to the values, which make up  the discrete spectrum of free H. Moreover, they approach these values  exponentially fast, since their asymptotics is created  by approaching the
argument of the factor $\G^{-1}(b)$, entering the asymptotics  of
the Kummer function $\F (b,c,z)$, to the pole $b \to -n_r, \
n_r=0,1,\dots.$  It would be worth to emphasize here once more that for atomic H in the cavity with the Robin boundary condition the Runge-Lenz vector is no longer conserved \cite{AlHashimi2012a,AlHashimi2012b}, therefore  these levels should be labeled now with two quantum numbers $n=n_r+1$ and $l$.

In particular, for the ``normal'' $ns$-levels one finds
\begin{multline} \label{f29}
E_{n0}(R) - E_{n0} \to \\  { \l -\g_{n0} \over \l +\g_{n0}} \ \[ {\g_{n0} \over n! } \]^2 \  \( 2 \g_{n0} R\)^{2n} \ e^{-2\g_{n0} R} \ , \quad  \g_{n0} R \gg 1 \ ,
\end{multline}
where
 \beq \label{f30}
E_{n0}= - \g_{n0}^2 / 2 \ , \quad \g_{n0} = q / n  \ , \quad n=1,2,\dots \ ,
\eeq
are the electronic $ns$-levels of the free atom. Remark, that levels with $\g_{n0} < \l$ should approach their asymptotics from above, while those with $\g_{n0} > \l$ from below.

If  $\l = \pm \g_{n0}$, the asymptotics (\ref{f29})  modifies in the
next way. The exponential behavior is preserved, while
the non-exponential factor undergoes changes in such a way, that
the levels approach their asymptotics of the free atom only from above. For  $\l =\g_{n0}$  their asymptotics takes the form
\begin{multline} \label{f31}
E_{n0}(R) - E_{n0} \to (n-1) \[ {\g_{n0} \over n! } \]^2  \ \( 2 \g_{n0} R\)^{2(n-1)} \ e^{-2\g_{n0} R} \ , \\ \g_{n0} R \gg 1 \ ,
\end{multline}
while for the lowest level  $E_{10}(R)$ the exponential part disappears completely, since in this case $\l=\g_{10}=q$ and   $E_{10}(R)=E_{1s}=-q^2/2$.

For $\l =-\g_{n0} < 0$ instead of (\ref{f29}) one obtains
\begin{multline} \label{f32}
E_{n0}(R) - E_{n0} \to \\  { 1 \over n+1} \  \[ {\g_{n0} \over n! } \]^2  \ \( 2 \g_{n0} R\)^{2(n+1)} \ e^{-2\g_{n0} R} \ , \quad \g_{n0} R \gg 1 \ .
\end{multline}
Moreover, in this case the limiting point  of the level $\tE_0 (R)$ with the power
asymptotics (\ref{f27})  coincides with the corresponding level  of the
free atom (\ref{f30}), that in turn represents a remarkable example
of von Neumann-Wigner avoided crossing effect, i.e. near levels reflection  under perturbation \cite{Neumann1929a,Neumann1929b,LandauLifshitz}
--- infinitely close to each other for $R \to \inf$ levels
$E_{n0}(R)$ and $\tE_0 (R)$ should for decreasing $R$ diverge in
opposite directions from their common limiting point  $E_{n0}$.
As a perturbation here serves the  nucleus
Coulomb field, since  under Robin  condition  the electronic WF doesn't vanish on the
cavity border, and so  for $R \gg a_B$ the maximum of electronic
density shifts to the region of large distances
between the electron and nucleus, where the contribution of the
Coulomb field is negligibly small compared to boundary effects. When
$R$ decreases, the Coulomb field increases, hence,   $E_{n0}(R)$
should go upwards according to (\ref{f32}), while $\tE_0 (R)$ goes
downwards according to asymptotics
\beq \label{f33} \tE_0 (R) \to
E_{n0} -  {n+1 \over n} \ {q \over R} + O(1/R^2) \ , \ R \to \inf \ .
\eeq

As a result, in a spherical cavity under   Robin condition and  nucleus in the center: i) for $\l=q$ the lowest  level of the one-electron atom $E_0(R)$ acquires for any $R$ the constant value $E_{1s}$ of the free atom; ii) for $\l > -q $ and $R \gg a_B$ it approaches $E_{1s}$ exponentially fast; iii) for $\l \leq -q <0$ it transforms into the level  $\tE_0(R)$ with the power-like asymptotics (\ref{f27}), which  in the whole range of $R$   behaves similar to a hyperbole \cite{Sveshnikov2013c, Sveshnikov2013d}.

Proceeding further, we'll see that all the above-mentioned effects, even  the last one, one way or another manifest themselves for atomic H over plane.

\section{Exactly solvable cases for atomic H over plane}\label{sect:2}

For atomic H over plane there exist three partial cases, when the corresponding spectral problem   (\ref{eq:1_02}), (\ref{eq:1_03}) allows for either exact or  ``quasi-exact'' analytic solution. The first one is the already mentioned above result \cite{Pupyshev2000,Sen2009,Sveshnikov2013a,Sveshnikov2013b,Sveshnikov2013c,Sveshnikov2013d}, well-known in  quantum chemistry, that if the nucleus charge  $q$ and the surface interaction constant are related via $\lambda=q$, then for such one-electron atom in a spherical cavity of radius $R$ for any $0 < R \leq \infty$ the lowest  $s$-level is given by the exact solution of SE for the $1s$ state of the free atom, that means
\begin{equation}\label{eq:2_00}
E_{10}(R) = E_{1s}=-q^2/2 \  .
\end{equation}
The nucleus in this case resides exactly in the cavity center without displacement. So by passing to the limit $R \to \infty$ one obtains the exact solution for the lowest eigenstate of atomic H over plane with the coupling constant $\lambda=1$, which implies that the atom has pushed off from the plane to infinity. In turn, it means that for $\lambda \geq 1$ the mutual reflection between  H and plane should be so strong, that the minimal energy can be reached only for infinite removal of H from the plane.

Two other  cases, which allow for a ``quasi-exact'' solution, correspond to Dirichlet ($\lambda\rightarrow +\infty$) and Neumann ($\lambda =0$) boundary conditions.
For these purposes let us pass to  prolate spheroidal coordinates $(\xi, \eta, \varphi) \ , \ 1 \leq \xi \leq \infty \ , \ -1 \leq \eta \leq 1 \ , \ 0 \leq \varphi < 2 \pi $, in which the nucleus resides  in the focus with coordinates $\xi = 1$ and $\eta = 1$, while the boundary plane $z=0$ is described  by the condition $\eta =0$. In this coordinate frame the Dirichlet and Neumann boundary conditions take the form
\begin{equation}\label{eq:2_01}
\psi\Big|_{\eta=0}=0 \quad \text{and}\quad \dfrac{\partial}{\partial \eta}\psi\Big|_{\eta=0}=0 \ ,
\end{equation}
respectively, while SE allows for separation of variables \cite{Komarov1976}.

Implying that $\psi=\Pi(\xi)\,\Xi(\eta)\,e^{\pm i m\varphi}$, from (\ref{eq:1_02}) one obtains
\begin{equation}\label{eq:2_02}
\begin{aligned}
&\frac{d}{d\xi}(\xi^2-1)\frac{d}{d\xi}\Pi+\left[-\mu-p^2(\xi^2-1)+a\xi-\frac{m^2}{\xi^2-1}\right]\Pi=0 \ , \\
&\frac{d}{d\eta}(1-\eta^2)\frac{d}{d\eta}\Xi+\left[\mu-p^2(1-\eta^2)+b\eta-\frac{m^2}{1-\eta^2}\right]\Xi=0 \ ,
\end{aligned}
\end{equation}
where $\mu$ is the separation parameter, $p^2=-2Eh^2$ and $a=b=2h$. The corresponding eigenfunctions are seeked in the form of the following power series \cite{Komarov1976}
\begin{equation}\label{eq:2_03}
\begin{aligned}
\Pi(\x)&=(\xi^2-1)^{m/2}e^{-p(\xi-1)}(\xi+1)^\sigma\sum_{s=0}^{\infty}g_s x^s \ ,\\
\Xi(\eta)&=(1-\eta^2)^{m/2}e^{-p(1-\eta)}\sum_{s=0}^{\infty}c_s (1-\eta)^s \ ,
\end{aligned}
\end{equation}
where $\sigma=1/\sqrt{-2E}-(m+1)$, while $x=(\xi-1)/(\xi+1)$. The expansion coefficients $g_s$ and $c_s$ are subject of the  three-term recurrence relations
\begin{equation}\label{eq:2_04}
\begin{aligned}
\omega_s g_{s+1}-\tau_s g_s+\gamma_s g_{s-1}=0 \ ,\\
\rho_s c_{s+1}-\kappa_s c_s+\delta_s c_{s-1}=0 \ ,
\end{aligned}
\end{equation}
wherein $g_{-1}=c_{-1}=0$, while
\begin{equation}\label{eq:2_05}
\begin{aligned}
&\left\{\begin{aligned}
&\omega_s=(s+1)(s+m+1)\ , \\
&\tau_s=2s(s+2p-\sigma)-(m+\sigma)(m+1)-2p\sigma+\mu \ ,\\
&\gamma_s=(s-1-\sigma)(s-m-1-\sigma) \ ,\\
\end{aligned}\right.\\
&\left\{\begin{aligned}
&\rho_s=2(s+1)(s+m+1) \ ,\\
&\kappa_s=s(s+1)+(2s+m+1)(2p+m)-b-\mu \ ,\\
&\delta_s=2p(s+m)-b \ .
\end{aligned}\right.
\end{aligned}
\end{equation}
The conditions of regularity for $\Pi(\xi)$ on the interval $1 \leq \xi \leq \infty$ and $\Pi(\xi)\rightarrow 0$ for $\xi\rightarrow\infty$ lead to relation
\begin{equation}\label{eq:2_06}
\begin{Vmatrix}
-\tau_0 & \omega_0 & 0 & 0 & \cdots \\
\gamma_1 & -\tau_1 & \omega_1 & 0 & \cdots \\
0 & \gamma_2 & -\tau_2 & \omega_2 & \cdots  \\
\hdotsfor{5}
\end{Vmatrix}=0\; ,
\end{equation}
while the Dirichlet and Neumann boundary conditions (\ref{eq:2_01}) take the form
\begin{equation}\label{eq:2_07}
\Xi(\eta){\vrule height 2 ex depth 2 ex }_{\; \eta=0}=0 \Leftrightarrow \sum_{s=0}^{\infty}c_s =0 \ ,
\end{equation}
\begin{equation}\label{eq:2_08}
\frac{\partial}{\partial\eta}\Xi(\eta){\vrule height 2 ex depth 2 ex }_{\; \eta=0}=0 \Leftrightarrow \sum_{s=0}^{\infty}(p-s)c_s =0 \ ,
\end{equation}
respectively.
\begin{figure}[ht!]
		\includegraphics[width=\columnwidth]{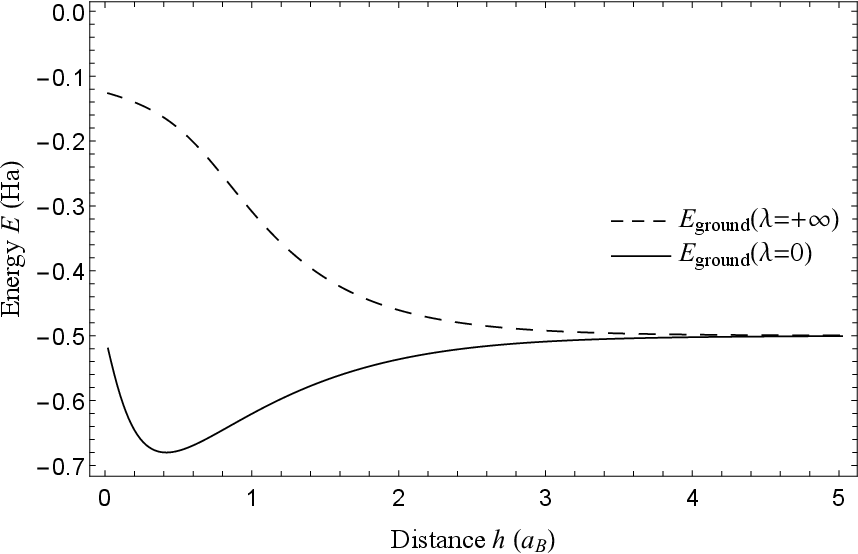}
\caption{Atomic H ground state energy $E(h)$    for the Dirichlet (dashed line) and Neumann (solid line) conditions. }\label{img:01}
\end{figure}

So  upon cutting the series from above at certain finite $s_{max}$ the search for the energy levels reduces to solution of the systems of algebraic equations  (\ref{eq:2_06}), (\ref{eq:2_07}) and (\ref{eq:2_06}), (\ref{eq:2_08}) with respect to $\mu$ and $E$. And since the series  (\ref{eq:2_03}) are rapidly convergent \cite{Komarov1976}, it suffices to take $s_{max} \simeq 10-20$ to provide  the required precision of calculations (see Section \ref{sect:4}).

\begin{figure}[hb!]
		\includegraphics[width=\columnwidth]{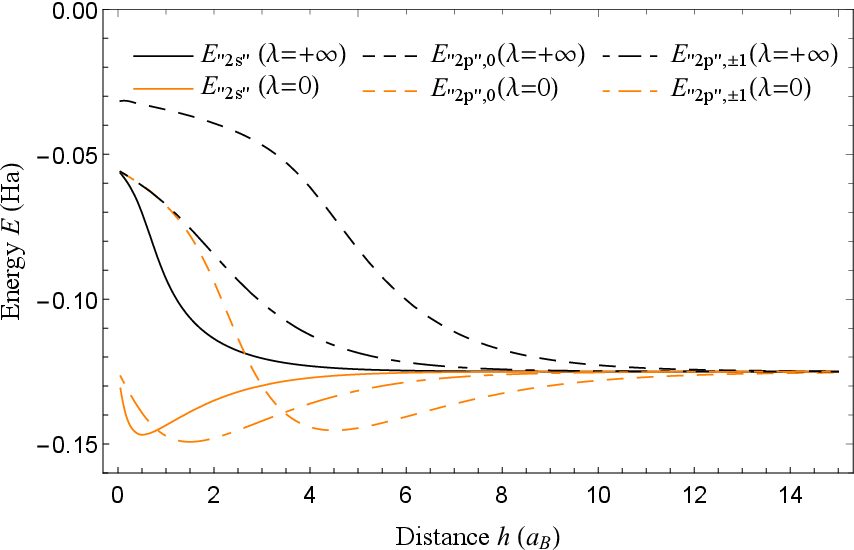}
\caption{(Color online). Atomic H energy $E(h)$ of two first excited levels ``$2s$'' and ``$2p$''.	 For $h \gg a_B$ the solid curves tend to  $2s$-level of the free atom, the dashed ones  -- to $2p,m=0$, while the  dash-dotted ones  -- to $2p, m=\pm 1$, respectively, with the common asymptotics of the bound energy, corresponding to $n=2$ of the free H. Black  curves correspond to the Dirichlet condition, while the orange curves  -- to the Neumann one. }\label{img:02}
\end{figure}

The results of such ``quasi-exact'' analytic calculations with the Dirichlet and Neumann boundary
conditions for the ground state and the first excited levels ``$2s$'' and ``$2p$''are shown in Figs.
\ref{img:01}-\ref{img:02}, whence it follows that for atomic H over plane the  behavior of levels
turns out to be sufficiently different depending on the type of boundary conditions.
Under the Dirichlet condition the atom repels itself from the plane, whereas under
the Neumann one the atom will occupy an equilibrium position at some finite distance
from the plane (see Fig.\ref{img:01}). The latter effect is quite understandable,
since the Dirichlet condition appears in the limit $\lambda\rightarrow +\infty$,
i.e. when the atom reflects from the plane with the maximal force, whereas the
Neumann one corresponds to the case of the neutral boundary $\lambda =0$.
At the same time, in both cases with increasing $h$ the ground state levels tend
exponentially fast to their common asymptotics, corresponding to that of the free H.  Here it would be worthwhile to emphasize that in the  considered picture  the medium, filling the half-space $z<0$,  is modeled by the boundary condition (\ref{eq:1_03}), which represents the summary of all the excitations in the bulk and on its surface and so can be either attractive or repulsive. Therefore, there is every reason for the resulting  asymptotics of levels to be significantly different from  the attractive van der Waals potential $V_A(h)$, which should fall off asymptotically as $\sim 1/h^6$, and all the more from various corrections like the Casimir-Polder force \cite{Itzykson1980}.

Another picture appears for $h\rightarrow 0$. For the Dirichlet condition the electronic WF of atomic H, which nucleus resides directly on the boundary surface $z=0$, should be odd with respect to reflection in the plane and so admits only those  $nlm$-atomic levels, for which $l+m$ are odd. In the case of the Neumann condition WF should be even under reflection, hence, the values of  $l+m$ should be even too.
Therefore by transition from $h \to \infty$ to $h \to 0$ the  $1s$-level transforms into $2p$ with $m=0$ for the Dirichlet condition and into $1s$ for the Neumann one. Remark that during this transition the third quantum number remains unchanged, since $l_z$ is conserved.

A similar picture takes place also for the excited levels ``$2s$'' and ``$2p$'' (see Fig.\ref{img:02}).
In particular,  in the case of Dirichlet condition the calculations show that by transition from  $h \to \infty$ to $h \to 0$ they don't intersect. Therefore each level corresponds to its  ``unique'' local extremum of the energy functional, fixed in relation to others. It follows whence that during such transition the level $2s$ should transform into $3p$ with $m=0$, which is the next allowed for $h \to 0$ after $2p$ with $m=0$, which in turn is the result of $1s$-transmutation. At the same time, the level $2p$ with $m=0$ of the free atom for $h \to \infty$, being allowed for $h \to 0$ too, actually rises even higher up to $4f$ with $m=0$. Here the repulsive property of the Dirichlet boundary condition shows up very clearly. Finally, the level $2p$ with $m=\pm 1$ for $h \to \inf$  transforms into allowed for $h \to 0$ level $3d$ with $m=\pm 1$. Such a specific splitting of the first excited ($n=2$) energy level of free  H by transition from $h \to \infty$ to $h \to 0$ is closely related to the spherical symmetry breakdown in the problem considered. In the case of $\Re^3/2$ with plane boundary there remains instead of the spherical symmetry only the axial one, hence, for finite $h$ the orbital moment isn't conserved any more, there remains only its projection  $l_z$. As a result, for finite  $h$ the electronic WF is a  superposition of a large amount of spherical harmonics with their specific radial components, whose analytic form can be found only in the partial cases considered above.

In the Neumann case  all the lowest excited levels ``$2s$'' and ``$2p$'' with $m=0\, , m=\pm 1$ intersect each other, that follows directly both from the results of ``quasi-exact''  analytic solution (\ref{eq:2_06}), (\ref{eq:2_08}), shown in Fig.\ref{img:02}, and from the variational estimates based on the special type trial functions, which  are explored below in Sect.\ref{sect:3}. In this case by transition from $h \to \infty$ to $h \to 0$ the level $2s$ transforms into  $2s$, $2p$ with $m=0$ -- into $3d$ with $m=0$, $2p$ with $m=\pm 1$ -- into  $2p$ with $m= \pm 1$, and everytime the energy minimum  takes place for some finite but non-zero  $h$. The latter circumstance reflects the fact that under Neumann condition the direct interaction between the atomic electron and the boundary is actually absent, while the boundary condition itself provides only the electron's ``not going through'' property into the forbidden  region $z<0$. Therefore it turns out energetically favorable for the atom to be  in the mode of ``soaring'' over the boundary plane  at some finite height.

\section{Variational estimates for the lowest levels of atomic H over plane with $\lambda \gtrsim - 0.3$}\label{sect:3}

Now -- having dealt with the exactly solvable cases this way -- let us turn to  a more general situation, when the electronic WF on the plane is subject of Robin condition (\ref{eq:1_03}) with $\lambda(\vec{\rho})=const \not= 0\, , 1, \infty$. In this case the analytic solution of SE is absent, and so the variational estimates and numerical tools come into play.

As it was shown in the preceding Section, for $\lambda \geq 1$ the mutual reflection between  H and plane should be so strong, that the minimal energy is reached only in the case of infinite separation  between  them. At the same time, for  $\lambda<1$ the effective atomic potential becomes attractive, hence, the minimal energy of the ground state can be reached at finite distances from the boundary surface. This picture is confirmed both via direct numerical calculations (whose detailed description is presented in the next Section \ref{sect:4}) and by means of variational estimates with special type electronic trial WF.

In this Section we consider  the variational estimates, based on a special modification of the effective charge approximation, which works  for $\l \gtrsim - 0.3$. Since the problem is axially symmetric, we'll use the cylindric coordinate frame $(\rho,\varphi, z)$, in which the nucleus is placed at the point  $(0,0,h)$.
The exact solution for energy levels corresponds to the local extremes of the functional
\begin{multline}\label{eq:3_01}
E[\psi]=\frac{1}{N}
\Bigg[ \int\limits_{z \geq 0} dz \,\rho\, d\rho \  \(\frac{1}{2}|\vec\nabla\psi|^2 -
\frac{|\psi|^2}{\sqrt{\rho^2+(z-h)^2}}\)  \ + \\ + \ \frac{\lambda}{2}\,\int\limits_{z =0} \rho\, d\rho\,|\psi|^2 \Bigg] \ ,
\end{multline}
\begin{equation}\label{eq:3_02}
N=\int\limits_{z\geq 0} dz \,\rho \,d\rho\, |\psi|^2\; .
\end{equation}

The choice of trial functions is based mainly on the following argument. For such values of $\l$ in presence of the plane the behavior of the electronic WF in the vicinity of the nucleus cannot significantly change, especially for sufficiently large separation between the nucleus and plane. The main effect of interaction between the atomic electron and the border reduces in this case to the screening of the nucleus charge. Therefore it is natural to employ as a first approximation the exact WF of the Coulomb problem in the unbounded space, modified by an effective nucleus charge $q=2 \alpha$ combined with a correction of its behavior near the  plane $z=0$, where the boundary condition (\ref{eq:1_03}) should hold. Such a correction can be quite effectively implemented by a multiplier, which changes the logarithmic derivative of WF with respect to $z$-coordinate.  The most suitable version of such a multiplier turns out to be $\exp(-\beta z)$. From one side, inserting  this factor enlarges the amount of analytic calculations for the mean energy, but from the other, it remarkably improves the agreement with  results of direct numerical calculations.

With account for these considerations the trial ground state WF will be taken in the form of $1s$-hydrogen function with an effective nucleus charge $q=2 \alpha$ and the additional multiplier $\exp(-\beta z)$, while for the first excited states with $m=\pm 1$ ---  $2p$-hydrogen functions with $m=\pm 1$ and the same  $q=2 \alpha$ and  additional multiplier (the choice of  trial WF for the first excited states ``$2s$'' and ``$2p$'' with $m=0$ is discussed below)
\begin{equation}\label{eq:3_03}
\psi^{tr}_{``1s"}=\exp(-2\alpha\sqrt{\rho^2+(z-h)^2}-\beta z) \ ,
\end{equation}
\begin{equation}\label{eq:3_04}
\psi^{tr}_{``2p",\pm 1}=\rho\exp(-\alpha\sqrt{\rho^2+(z-h)^2}-\beta z\pm i \varphi) \ .
\end{equation}
The appearance of the factor  $\rho$ in  $\psi^{tr}_{``2p",\pm 1}$ is caused by the fact that the complete  $2p$-hydrogen WF with $m=\pm 1$ possesses the angular part, containing $P_1^{\pm 1}(\cos \theta)=\pm \sin \theta$, and so the factor $r$ in the radial part of $2p$-hydrogen function transforms into $\rho$.

As a result, upon substituting (\ref{eq:3_03}) into (\ref{eq:3_01}) we obtain the following expression for the mean value of the ground state energy
\begin{equation}\label{eq:3_05}
E^{tr}_{``1s"}=\frac{A_0+(A_1+\lambda A_2)\exp{[-2 h(2\alpha+\beta)]}}{A_3+A_4\exp{[-2 h(2\alpha+\beta)]}} \ ,
\end{equation}
with the coefficients in the nominator being equal to
\begin{equation}\begin{gathered}\label{eq:3_06}
A_0=32\alpha^2\frac{\alpha-1}{8\alpha^2-2\beta^2} \ ,
\\ A_1=-\frac{\beta}{2}-\frac{4\alpha(\alpha-1)}{2\alpha+\beta}-2h\alpha(2\alpha+\beta) \ ,
\\ A_2=1+2h\alpha \ ,
\end{gathered}\end{equation}
while in the denominator
\begin{equation}\label{eq:3_07}
A_3=\frac{32\alpha^3}{(4\alpha^2-\beta^2)^2} \ , \quad A_4=\frac{-\beta-4\alpha(1+h(2\alpha+\beta))}{(2\alpha+\beta)^2} \ .
\end{equation}
For the first excited ``$2p$''-state with $m=\pm 1$ the mean value upon substituting (\ref{eq:3_04}) into (\ref{eq:3_01}) takes the form
\begin{equation}\label{eq:3_08}
E^{tr}_{``2p", \pm 1}=\frac{B_0+(B_1+\lambda B_2)\exp{[-2 h(2\alpha+\beta)]}}{B_3+B_4\exp{[-2 h(2\alpha+\beta)]}} \ ,
\end{equation}
with coefficients
\begin{equation}\begin{gathered}\label{eq:3_09}
B_0=\frac{\alpha-1}{2(\alpha-\beta)^2(\alpha+\beta)^2} \ ,\\
B_1= -\Big(-8\alpha^2+8\alpha^3-4\alpha\beta+7\alpha^2\beta \ + \\ + \ 6\alpha\beta^2+3 \beta^3+4 h^2 \alpha^2(\alpha+\beta)^3 + \  2 h \alpha (\alpha+\beta) (5\alpha^2 \ +
\\ + \ 4\alpha (-1+\beta)+3\beta^2\Big) /  \Big(32\alpha^4 (\alpha+\beta)^2\Big) \ , \\
B_2= \frac{3+6 h\alpha + 4 h^2\alpha^2}{16\alpha^4} \ ,
\end{gathered}\end{equation}
and
\begin{equation}\begin{gathered}\label{eq:3_10}
 B_3=\frac{\alpha}{(\alpha^2-\beta^2)^3} \ ,\\
B_4= - \Big(2\alpha^2(4+h\alpha(5+2 h \alpha))+\alpha (9+8 h \alpha(2+h \alpha)) \beta \ + \\ + \ (3+2 h\alpha(3+2 h \alpha)) \beta^2\Big)/\Big(16\alpha^4 (\alpha+\beta)^3\Big) \ .
\end{gathered}\end{equation}
Here it should be noted that the condition of orthogonality to the ground state for the  trial functions of excited levels is exactly fulfilled only for the trial WF (\ref{eq:3_04}) with $m=\pm 1$.
In the trial WF for excited levels with $m=0$ there might exist an admixture of the exact ground state WF. Therefore the variational estimate for the ground state majorizes always  its exact energy  from above, the same should take place for the first excited state with $m=\pm 1$, whereas for the excited states with $m=0$ in the general case this statement might be incorrect.

So for  two remaining first excited states ``$2s$'' and  ``$2p$'' with $m=0$ the choice of the trial functions turns out to be  more complicated. First, there follows from the analytic solution in the Neumann case (see Sect.\ref{sect:2}) that in presence of the boundary plane such states undergo hybridization. In this case the natural choice for trial functions is  their linear combination
\begin{equation}\label{eq:3_11}
\psi^{tr}_{exct,m=0}=\psi^{tr}_{``2s"}\cos(\chi)+\psi^{tr}_{``2p",0}\sin(\chi) \ ,
\end{equation}
where $\chi$ is the mixture  variational parameter,
\begin{equation}\label{eq:3_12}
\psi^{tr}_{``2s"}=(\sqrt{\rho^2+(z-h)^2}-\gamma)\exp(-\alpha\sqrt{\rho^2+(z-h)^2}-\beta z) \ ,
\end{equation}
\begin{equation}\label{eq:3_13}
\psi^{tr}_{``2p",0}=(z-h)\,\exp(-\alpha\sqrt{\rho^2+(z-h)^2}-\beta z) \ .
\end{equation}
The multiplier $(z-h)$ in  $\psi^{tr}_{``2p",0}$ is caused by the same reason as $\rho$ in $\psi^{tr}_{``2p",\pm 1}$.

The parameter  $\gamma$, which enters the expression (\ref{eq:3_12}) for $\psi^{tr}_{``2s"}$, is determined from the condition of orthogonality between $\psi^{tr}_{exct,m=0}$ and the trial ground state function $\psi^{tr}_{``1s"}$  (\ref{eq:3_03})
\begin{equation}\label{eq:3_14}
\int\limits_{z\geq 0}dz\, \rho\, d\rho\, \psi^{tr}_{``1s"}\, \psi^{tr}_{exct,m=0}=0 \ ,
\end{equation}
in order to reduce, if possible,   the admixture of the ground state WF (but not to remove completely!). The explicit answer for $\gamma$ reads
\begin{multline}\label{eq:3_15}
\gamma= \\ \dfrac{\dfrac{C_0}{(9\alpha^2-4\beta^2)^3}+\dfrac{\exp[-3h\alpha-2h\beta]}{(27\alpha^3(3\alpha+2\beta)^3)} (C_1\sin(\chi) + C_2\cos(\chi))}{\dfrac{C_3}{(9\alpha^2-4\beta^2)^2} +\dfrac{\exp[-3h\alpha-2h\beta]}{9 \alpha^2 (3 \alpha + 2 \beta)^2}C_4 \cos(\chi)} \ ,
\end{multline}
where the coefficients in the nominator are equal to
\begin{multline}\label{eq:3_16}
C_0=-4\Big((27\alpha^2 + 4\beta^2)\cos(\chi) -24\alpha\beta \sin(\chi)\Big) \ , \\
C_1=3\alpha \Big(9\alpha+27h\alpha^2 +27h^2\alpha^3 +2\beta +24h\alpha\beta \ + \\ + \  36h^2\alpha^2\beta + 4h\beta^2 +12h^2\alpha \beta^2\Big) \ ,\\
C_2=54\alpha^2 +108h\alpha^3 +81h^2\alpha^4 +36\alpha\beta +108h\alpha^2\beta \ +\\
+ \ 108h^2\alpha^3 \beta +8\beta^2 +24h\alpha \beta^2 +36h^2\alpha^2 \beta^2 \ ,
\end{multline}
while in the denominator
\begin{equation}\label{eq:3_17}
C_3=-12 \alpha \cos(\chi) \ , \quad
C_4=6 \alpha +9h\alpha^2 + 2\beta +6h\alpha \beta \ .
\end{equation}

Upon substituting the representation (\ref{eq:3_11}) for $\psi^{tr}_{exct,m=0}$ into (\ref{eq:3_01}) the mean value of energy for these levels takes the form
\begin{equation}\label{eq:3_18}
E^{tr}_{exct,m=0}=\frac{N_0+(N_1+\lambda N_2)\exp[-2h(\alpha+\beta)]}{N_3+N_4\exp[-2h(\alpha+\beta)]} \ .
\end{equation}
The coefficients $N_i$ in eq.(\ref{eq:3_18}) have the following structure
\begin{equation}\label{eq:3_19}
N_i=\frac{a_i+b_i\sin(2\chi)+c_i\cos(2\chi)}{d_i} \ ,
\end{equation}
where\begin{widetext}
\begin{equation}\label{eq:3_20}
\left\{\begin{aligned}
&a_0=-4\alpha^2+2\alpha^3-4\beta^2+2\alpha\beta^2+4\alpha^3\gamma-\alpha^4\gamma -4\alpha\beta^2\gamma+\beta^4\gamma-2\alpha^4\gamma^2+\alpha^5\gamma^2+\\
&+4\alpha^2 \beta^2\gamma^2-2\alpha^3\beta^2\gamma^2-2\beta^4\gamma^2+\alpha\beta^4\gamma^2 \ , \\
&b_0=2 (\alpha-2) \beta ( \alpha^2 \gamma-\beta^2 \gamma-2\alpha) \ , \\
&c_0=(\alpha^2-\beta^2) (\alpha^3\gamma^2+\beta^2\gamma (2\gamma-1)-\alpha^2\gamma (1+2\gamma)+\alpha\gamma(4-\beta^2\gamma)-2) \ , \\
&d_0=8(\alpha^2 - \beta^2)^3\ ,
\end{aligned}\right.
\end{equation}
\begin{equation}\label{eq:3_21}
\left\{\begin{aligned}
&a_1=-16\alpha^3+8\alpha^4-24h\alpha^4+10h\alpha^5-16h^2\alpha^5+4h^2\alpha^6+8h^3\alpha^7-12\alpha^2\beta+9\alpha^3\beta-\\
&-32h\alpha^3\beta+16h\alpha^4\beta-
32h^2\alpha^4\beta+12h^2\alpha^5\beta+32h^3\alpha^6\beta-4\alpha\beta^2+
11\alpha^2\beta^2-8h\alpha^2\beta^2+\\
&+16h\alpha^3\beta^2-16h^2\alpha^3\beta^2+
20h^2\alpha^4\beta^2+48h^3\alpha^5\beta^2+9\alpha\beta^3+16h\alpha^2\beta^3+
20h^2\alpha^3\beta^3+\\
&+32h^3\alpha^4\beta^3+3\beta^4+6h\alpha\beta^4+8h^2\alpha^2\beta^4+8h^3\alpha^3\beta^4+16\alpha^4\gamma-4\alpha^5\gamma+16h\alpha^5\gamma-\\
&-8h\alpha^6\gamma-8h^2\alpha^7\gamma+24\alpha^3\beta\gamma-8\alpha^4\beta\gamma+ 32h\alpha^4\beta\gamma-24h\alpha^5\beta\gamma-32h^2\alpha^6\beta\gamma+\\
&+
8\alpha^2\beta^2\gamma-12\alpha^3\beta^2\gamma+16h\alpha^3\beta^2\gamma-32h\alpha^4\beta^2\gamma-48h^2\alpha^5\beta^2\gamma-12\alpha^2\beta^3\gamma-\\
&-24h\alpha^3\beta^3\gamma-32h^2\alpha^4\beta^3\gamma-4\alpha\beta^4\gamma-8h\alpha^2\beta^4\gamma-8h^2\alpha^3\beta^4\gamma-8\alpha^5\gamma^2+4\alpha^6\gamma^2+\\
&+4h\alpha^7\gamma^2-16\alpha^4\beta\gamma^2+10\alpha^5\beta\gamma^2+16h\alpha^6\beta\gamma^2-8\alpha^3\beta^2\gamma^2+10\alpha^4\beta^2\gamma^2+24h\alpha^5\beta^2\gamma^2+\\
&+6\alpha^3\beta^3\gamma^2+16h\alpha^4\beta^3\gamma^2+2\alpha^2\beta^4\gamma^2+4h\alpha^3\beta^4\gamma^2 \ , \\
&b_1=2\alpha(\beta^3(2h\beta-1)+4h^2\alpha^6(h-\gamma)+2h\alpha^5(1+8h\beta)(h-\gamma)+\alpha\beta(2h\beta^3(2h-\gamma)+\\
&+\beta^2(4h+\gamma)-2-(3+4h)\beta)+\alpha^4(24h^3\beta^2-\gamma+h(4-4(\beta-2)\gamma)+\\
&+h^2(6\beta-24\beta^2\gamma-8))+\alpha^3(2+16h^3\beta^3+(4+\beta)\gamma-2h^2\beta(8-5\beta+8\beta^2\gamma)-\\
&-4h(3+\beta^2\gamma-\beta(1+4\gamma)))+\alpha^2(2h\beta^3(5h-2\gamma)-2\beta(1+8h-2\gamma)+\\
&+4h^2\beta^4(h-\gamma)+\beta^2(-8h^2+3\gamma+h(2+8\gamma))-6)) \ , \\
&c_1=-(\alpha+\beta)(-3\beta^3+4h\alpha^6(2h-\gamma)\gamma+2\alpha\beta(2-3\beta+\beta^2(2\gamma-3 h))-\\
&-4\alpha^5(\gamma^2+h^2(1-6\beta\gamma)+h\gamma(3\beta\gamma-2))+\alpha^2(8+\beta(8h-8\gamma-3)+\\
&+\beta^2(8\gamma-14h)-2\beta^3(2h^2-4h\gamma+\gamma^2))+2\alpha^4(\gamma(2+4\gamma-3\beta\gamma)+6h^2\beta(2\beta\gamma-1)-\\
&-h(1+8\gamma-8\beta\gamma+6\beta^2\gamma^2))+2\alpha^3(2h^2\beta^2(2\beta\gamma-3)+2\gamma(\beta+2\beta\gamma-\beta^2\gamma-4)+\\
&+h(4-5\beta-8\beta\gamma+8\beta^2\gamma-2\beta^3\gamma^2))) \ , \\
&d_1=-64\alpha^4(\alpha+\beta)^3 \ ,
\end{aligned}\right.
\end{equation}
\begin{equation}\label{eq:3_22}
\left\{\begin{aligned}
&a_2=\frac{3}{2}+3h\alpha+4h^2\alpha^2+4h^3\alpha^3-2\alpha\gamma-4h\alpha^2\gamma-4h^2\alpha^3\gamma+\alpha^2\gamma^2+2h\alpha^3\gamma^2 \ , \\
&b_2=2h\alpha(1+\alpha(2h-\gamma)+2\alpha^2(h^2-h\gamma)) \ , \\
&c_2=\frac{3}{2}+3h\alpha+2h^2\alpha^2-2\alpha\gamma-4h\alpha^2\gamma-4h^2\alpha^3\gamma+\alpha^2\gamma^2+2h\alpha^3\gamma^2 \ , \\
&d_2=8\alpha^4 \ ,
\end{aligned}\right.
\end{equation}
\begin{equation}\label{eq:3_23}
\left\{\begin{aligned}
&a_3=4\alpha^3+8\alpha\beta^2-3\alpha^4\gamma+2\alpha^2\beta^2\gamma+\beta^4\gamma+\alpha^5\gamma^2-2\alpha^3\beta^2\gamma^2+\alpha\beta^4\gamma^2 \ ,\\
&b_3=-2\beta(5\alpha^2+\beta^2-2\alpha^3\gamma+2\alpha\beta^2\gamma) \  , \\
&c_3=(\alpha^2-\beta^2)(\alpha^3\gamma^2-3\alpha^2\gamma-\beta^2\gamma+\alpha(2-\beta^2\gamma^2)) \ , \\
&d_3=4(\alpha^2-\beta^2)^4 \ ,
\end{aligned}\right.
\end{equation}
\begin{equation}\label{eq:3_24}
\left\{\begin{aligned}
&a_4=16\alpha^3+26h\alpha^4+20h^2\alpha^5+8h^3\alpha^6+19\alpha^2\beta+46h\alpha^3\beta+48h^2\alpha^4\beta+24h^3\alpha^5\beta+\\
&+12\alpha\beta^2+26h\alpha^2\beta^2+36h^2\alpha^3\beta^2+24h^3\alpha^4\beta^2+3\beta^3+6h\alpha\beta^3+8h^2\alpha^2\beta^3+8h^3\alpha^3\beta^3-\\
&-12\alpha^4\gamma-16h\alpha^5\gamma-8h^2\alpha^6\gamma-24\alpha^3\beta\gamma-40h\alpha^4\beta\gamma-24h^2\alpha^5\beta\gamma-16\alpha^2\beta^2\gamma-\\
&-32h\alpha^3\beta^2\gamma-24h^2\alpha^4\beta^2\gamma-4\alpha\beta^3\gamma-8h\alpha^2\beta^3\gamma-8h^2\alpha^3\beta^3\gamma+4\alpha^5\gamma^2+4h\alpha^6\gamma^2+\\
&+10\alpha^4\beta\gamma^2+12h\alpha^5\beta\gamma^2+8\alpha^3\beta^2\gamma^2+12h\alpha^4\beta^2\gamma^2+2\alpha^2\beta^3\gamma^2+4h\alpha^3\beta^3\gamma^2 \ , \\
&b_4=2\alpha(\beta^2(1+2h\beta)+4h^2\alpha^5(h-\gamma)+\alpha\beta(1+2h\beta)(4+2h\beta-\beta\gamma)+\\
&+2h\alpha^4(6h^2\beta-3\gamma+h(5-6\beta\gamma))+2\alpha^2(1+2h\beta)(3+h^2\beta^2-2\beta\gamma+h\beta(4-\beta\gamma))+\\
&+\alpha^3(12h^3\beta^2-3\gamma-12h^2\beta(\beta \gamma-2)-2h(7\beta\gamma-6))) \ , \\
&c_4=-(\alpha+\beta)(4h\alpha^5(2h-\gamma)\gamma-3\beta^2+\alpha\beta(4\beta\gamma-9-6h\beta)-2\alpha^2(4+2h^2\beta^2-\\
&-6\beta\gamma+\beta^2\gamma^2-4h\beta(\beta\gamma-2))-4\alpha^4(\gamma^2+h^2(1-4\beta\gamma)+2h\gamma(\beta\gamma-2))+\\
&+2\alpha^3(4h^2\beta(\beta\gamma-1)-3\gamma(\beta\gamma-2)+h(12\beta\gamma-2\beta^2\gamma^2-5))) \ , \\
&d_4=-32\alpha^4(\alpha+\beta)^4 \ .
\end{aligned}\right.
\end{equation}\end{widetext}

\begin{figure}[ht!]
	\includegraphics[width=\columnwidth]{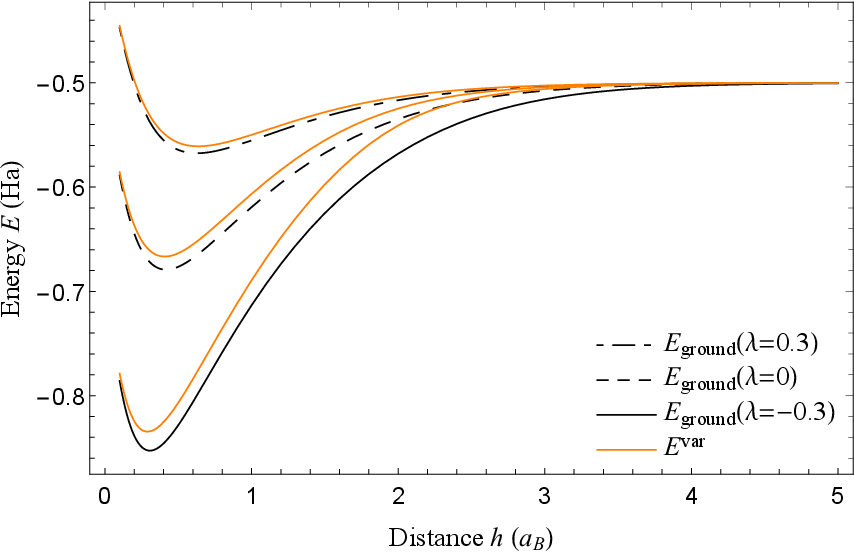}
	\caption{(Color online). Atomic H  ground state energy $E(h)$ as a function of the nucleus position over plane $h$ for the general Robin  condition on the boundary plane. Black curves correspond to direct numerical results, while  the orange ones -- to variational estimates by means of the trial function (\ref{eq:3_03}). }
	\label{img:03}
\end{figure}

\begin{figure}[hb!]
	\includegraphics[width=\columnwidth]{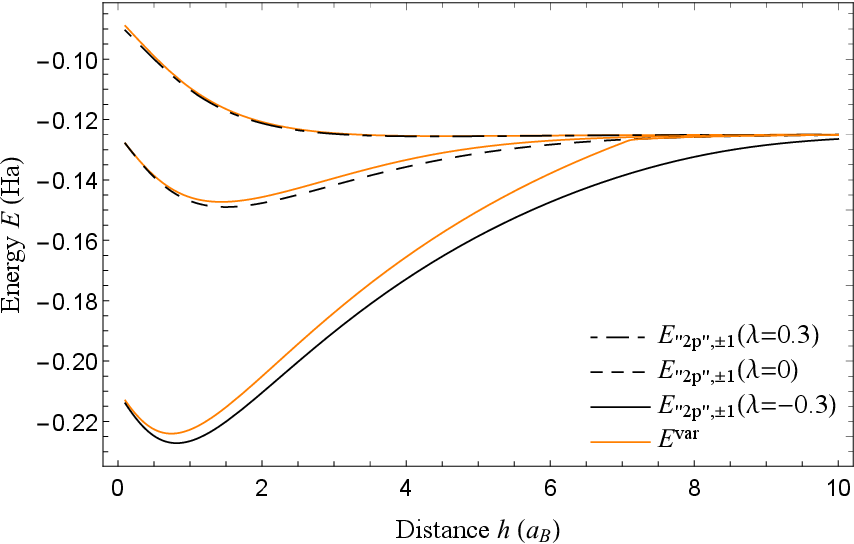}
	\caption{(Color online).  $E(h)$  of the first excited level with $m= \pm 1$, which tend to $2p$ for $h \gg a_B$, for the general Robin  condition on the boundary plane. Black curves correspond to direct numerical results, while the orange ones -- to variational estimates by means of the trial function (\ref{eq:3_04}).}
	\label{img:04}
\end{figure}

For $ \l \gtrsim - 0.3$ the minimization of expressions (\ref{eq:3_05}), (\ref{eq:3_08}) and (\ref{eq:3_18}) with respect to the variational parameters $\alpha\, , \beta$ gives a satisfactory agreement with the results of direct numerical calculations, at least  on the qualitative level. The dependence  of the ground state  ``$1s$'' and first excited ``$2p$'' with $m=\pm 1$  on the nucleus position $h$, found via such variational estimates, is shown in Figs.\ref{img:03},\ref{img:04}  in comparison with the direct numerical results.

Let us remark specially that by means of the trial functions (\ref{eq:3_03}),(\ref{eq:3_04}) and (\ref{eq:3_11})-(\ref{eq:3_13}) there are visible only the levels with exponential asymptotics for $h \gg a_B$, which in this case tend to the corresponding levels of free H with $n=1$ or $n=2$. At the same time, the levels with the power-like asymptotics, which  appear already for  $\lambda=-0.3$ and tend to the limiting point  $-\l^2/2=-0.045\, Ha$, cannot be caught  by means of such substitutions for $\psi^{tr}$, since their electronic WF possess a substantially different structure.

There follows from Fig.\ref{img:03} that for the considered values of $\lambda$ and $h \gtrsim 4\,a_B$ the ground state energy is practically indistinguishable from the $1s$ level of free H. On the other hand, with decreasing  $h$  the bound energy increases, and the faster, the more negative the value of $\lambda$, and hence, the stronger the attraction between the electron and the plane $z=0$. With decreasing $\lambda$ there also grows the maximal value of the bound energy, exceeding for $\lambda=0.3$ the value $0.568\,Ha$ at $h_{max}=0.627\,a_B$, for $\lambda=0$  --- $0.679\,Ha$ at $h_{max}=0.417\,a_B$, for $\lambda=-0.3$ --- $0.853\,Ha$ at $h_{max}=0.307\,a_B$. Collected  together, all these facts indicate that when the atom  approaches the plane, the electronic WF deforms the faster, the greater the attraction  between the atom and the boundary  plane.

It should be also noted that the difference between the variational estimates and direct calculations increases in the same way. Therefore, the  efficiency of the multiplier  $\exp(-\beta z)$, deforming the trial function in order to take account of the boundary plane $z=0$, decreases with varying  $\lambda$ from positive to negative values.

A similar  picture one finds in Fig.\ref{img:04}, where the dependence on $h$ for the level ``$2p$'' with $m=\pm 1$ in the same range  of $\lambda$ is presented. As in the case of the ground state, with increasing  $h$ all the curves tend to the corresponding energy $-0.125\,Ha$ of $2p$-level of free H. However, now  the levels closely approach their limiting value much later, at the scale  $h\gg10 a_B$. In addition, Fig.\ref{img:04} confirms explicitly the  remark concerning the exact orthogonality between the trial function (\ref{eq:3_04}) and the ground state WF, since the variational curves lie everywhere higher than the direct numerical results.

\begin{figure}[ht!]
\subfigure[]{
		\includegraphics[width=\columnwidth]{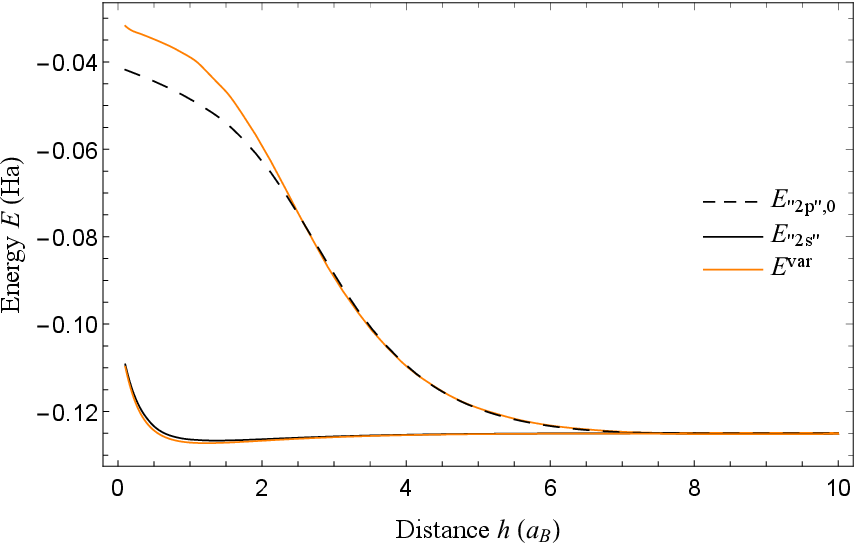}
}
\subfigure[]{
		\includegraphics[width=\columnwidth]{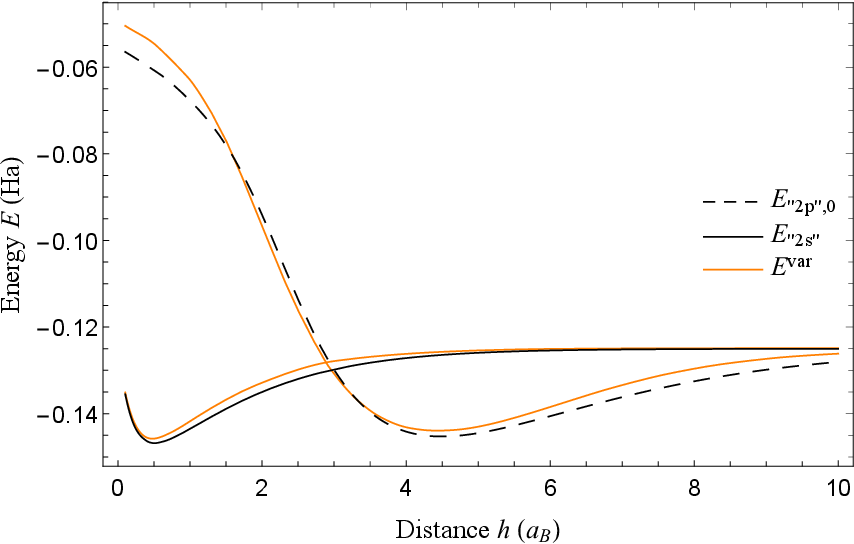}
}
\subfigure[]{
		\includegraphics[width=\columnwidth]{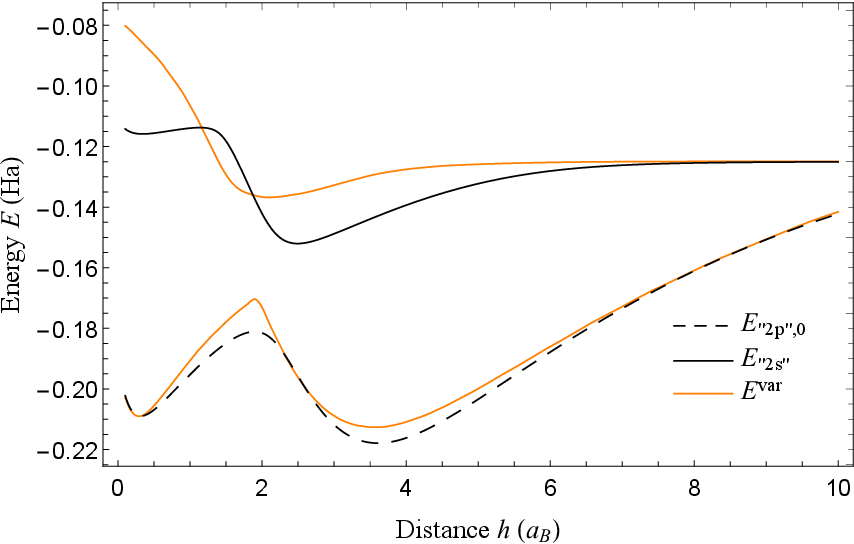}
}
\caption{(Color online). The dependence of two first excited levels with $m = 0$ on $h$ for (a): $\l= 0.3$; (b) $\l= 0$; (c) $\l=-0.3$. Black curves correspond to direct numerical results, while the orange ones -- to variational estimates by means of the trial function (\ref{eq:3_11})-(\ref{eq:3_13}). }
\label{img:05}	\end{figure}

In Figs.\ref{img:05} the results of variational estimates, based on the {\it Ansatz} (\ref{eq:3_11})-(\ref{eq:3_13}) for the trial function, and numerical calculations for the first two excited levels with $m=0$ and  $\lambda=0.3\, , 0\, , -0.3$, are presented. In this case the levels  óðîâíè approach closely their limiting values only at the scales $h\gg10 a_B$. From Fig.\ref{img:05}b there follows also that the trial function (\ref{eq:3_11}) permits to reproduce the hybridization of excited levels with  $m=0$,   obtained analytically in Sect.\ref{sect:2} for $\lambda=0$. It would be also worth to note that in this case  for certain values of $\l$ the variational curves lie below the numerical results, that points at  the presence of the ground state WF admixture in the trial functions (\ref{eq:3_11})-(\ref{eq:3_13}) and so in the mean value of the energy (\ref{eq:3_18}).

Thus, in the considered range of $\l$ the trial WF (\ref{eq:3_03}),(\ref{eq:3_04}) and (\ref{eq:3_11})-(\ref{eq:3_13}) are able to reproduce qualitatively  the behavior of corresponding atomic H energy levels, obtained via direct numerical calculations. As expected, the variational estimate reproduces the result the worse,  the more negative the value of $\lambda$. The reason is simple --- for sufficiently  negative $\l$  the true electronic WF transforms into  a superposition of a large amount of spherical harmonics in the form  of a  ``drop'' stuck and partially spread on the border plane, and hence, cannot be described in terms of such  simple trial functions.

\section{Atomic H over plane in the general case: direct numerical calculations and variational estimates for the lowest levels}\label{sect:4}

For atomic H in $\Re^3/2$ the most efficient numerical tool is the pertinent modification of the finite element method \cite{Ciarlet1978} within the direct variational approach to the functional (\ref{eq:3_01}), considered on a spatial lattice as a function of variables $\psi_{ij}=\psi(\rho_i,z_j)$. The calculations have been performed for the ground state and two first  excited states  in the range  $-1.2\leq\lambda\leq0.3$ and $0.2\,a_B\leq h\leq 10\,a_B$.
As the effective spatial infinity for the numerical problem $z_{max}=\rho_{max}=40\, a_B$ is chosen. The results of performed calculations confirm that such a choice  is quite satisfactory, since its further enlarging doesn't yield any appreciable changes within the precision achieved.  In particular, for the ground state and $h\simeq10\,a_B$ the relative error between the numerical and variational calculations grows up with decreasing $\l$, but doesn't exceed $\simeq1.5\times10^{-3}$. For the excited states in the considered range of $\lambda$ the relative error is $\sim 10^{-2}$.
Therefore the behavior of  the lowest levels  for $h \gtrsim 10\,a_B$  can be obtained without employing  cumbersome lattice calculations, just by means of variational estimates, considered below.

Within this  method  the integrals in the functional (\ref{eq:3_01}) are replaced by the integral sums according to trapezoid approximation, afterwards the extremes of the function of lattice variables $\psi_{ij}=\psi(\rho_i,z_j)$ are seeked. In the case of excited states the condition of orthogonality of corresponding solutions to the ground state WF is added.
The precision of calculations is controlled via changing the lattice step, that allows for additional increase of the accuracy of calculations by extrapolating the dependence of the obtained results on the magnitude of the square of the lattice spacing  to its value  tending to zero. In concrete calculations there have been used four subsequent lattices, the number of nodes in which  (per both coordinates) is related as 1:2:3:4.
The relative error, obtained by comparison the extrapolation result for three first lattices with the result of extrapolation for four lattices,  is  $O(10^{-4})$ for the ground state and $O(10^{-3})$ for the excited states.

Furthermore, the direct numerical analysis allows to trace how the levels tend to their asymptotic values for $h\rightarrow\infty$, since   for $\l<0$ the levels reveal two types of asymptotical behavior. If the electron-nucleus interaction  is stronger than the electron-plane one, then for $h \gg a_B$  the electron resides mostly in the vicinity of the nucleus, while its  bound energy tends to the corresponding level of the free H exponentially fast according to (\ref{f29}).

To the contrary, in the case of dominating interaction with plane the lowest electronic states correspond to the picture, wherein the electron leaves the nucleus and resides mostly in the vicinity of the plane, while the asymptotics of these levels with increasing $h$ is a power-like one similar to (\ref{f27}) with the common limiting point $(-\l^2/2)$. Quite similar to the case of an atom in the center of a large spherical cavity \cite{Sveshnikov2013c,Sveshnikov2013d}, this  effect becomes more pronounced the more negative   $\l$. It should be remarked, however, that in contrast to the case of a spherical cavity with $R \to \inf$, when the orbital moment is conserved and the power levels possess the quantum numbers  $lm$, in the presence of boundary plane even for $h \to \inf$ there remains only $l_z$ as a conserved quantity, and hence, these power-like levels can no longer be classified by the values of the orbital moment.

In the case under consideration for the ground state and the first excited levels one has two critical values of the coupling constant $\l_{crit,1}=-1$ and $\l_{crit,2}=-1/2$, respectively.
In the case when $\l> \lambda_{crit,2}$, the asymptotics of the ground state and the first excited ones coincides with the lowest levels $1s$, $2s$ and $2p$ of  free H with  energies $-1/2$ and $-1/8$, correspondingly, and this asymptotics is reached exponentially fast in the way similar to the law (\ref{f29}). When $\l= \lambda_{crit,2}$, the ground state asymptotics for $h\rightarrow\infty$ is still exponential and corresponds  to $1s$-level of free H, but the first excited ones transform into the second type of levels with power-like asymptotics and the common limiting point $- \lambda_{crit,2}^2/2=-1/8$, still coinciding  with the energy level with $n=2$ of free H. Moreover, as it was already mentioned in Sect.\ref{sect:1} (see formulae (\ref{f32}),(\ref{f33})), in this case due to  von Neumann-Wigner avoided crossing effect the power-like levels approach their limiting point $-1/8$ from below, whereas the exponential ones, corresponding to $2s$ and $2p$ levels of  free H, from above. When $\l_{crit,2}> \l > \l_{crit,1}$, the asymptotics of the ground state and of the first excited levels is $-1/2$ and $-1/2 < -\lambda^2/2<-1/8$, respectively, but the latter one is achieved remarkably slower, following a power law. In the  case  $\l = \l_{crit,1}$ there works once more the avoided crossing effect. Namely, the lowest level and a finite number of first excited reveal now the power-like behavior with the common limiting value equal to $E_{1s}$, while the exponential one becomes  the next excited level and approaches the same asymptotics from above. For details see Refs.\cite{Sveshnikov2013c}, Figs.7,8 and \cite{Sveshnikov2013d}, Fig.5. The only difference here is that in Refs.\cite{Sveshnikov2013c, Sveshnikov2013d} the nucleus is placed in the center of cavity and so both quantum numbers $lm$ serve as a superselection rule, whereas in the present case we have to deal outright  with a whole bundle of power-like excited levels without any definite value of the orbital moment. For $\l < \l_{crit,1}$  the  ground state together with the first excited levels tend to their common  limiting point $-\l^2/2<-1/2$ according to the power law, while all the others with growing $h$ approach exponentially fast the levels of free H. In this case we meet again the situation, when the exponential level, corresponding in the asymptotics to the $1s$ state of free H, becomes the first of the excited ones with exponential asymptotics, but not the first between all the excited levels, since  there will be a set of the first power-like excited levels,  which lie below $E_{1s}$.  It should be also noted that for $\l \leq \l_{crit,1}$  the atomic ground state represents a configuration, in which due to the strong attraction to the border its electronic  WF is  located almost completely in the vicinity of the latter with a small pick at the nucleus position, while the energy approaches its asymptotics for  $h\rightarrow\infty$  much slower, actually following the hyperbolic law.

As it was already mentioned in Sect.\ref{sect:3}, the trial functions (\ref{eq:3_03}),(\ref{eq:3_04}) and  (\ref{eq:3_11})-(\ref{eq:3_13}),  supplied with the effective nucleus charge and modulation of their behavior in the plane vicinity through the multiplier $\exp(-\beta z)$, cannot serve as the base for the estimates of power-like levels. For the approximate description of the latter some more complicated superpositions of hydrogen functions, modulated by the factors similar to $\exp(-\beta z)$ for an effective account of the boundary plane, are needed. For the first three  levels with $m=0$ it suffices to take into account only these multipliers, provided the linear combination  of the first 6 hydrogen functions is employed and the same argument as before in Sect.\ref{sect:3} is used: the behavior of the electronic WF in the nucleus vicinity cannot be strongly distorted in presence of the plane, especially for sufficiently large distances between the nucleus and plane.

Proceeding further this way, let us use as trial functions for three first states with $m=0$  the following superpositions
\begin{equation}\label{eq:4_01}
\psi^{tr}=\sum_{i=1}^6\,  c_i\,\psi_i^{tr} \ ,
\end{equation}
with $\psi^{tr}_i$ being the WF of first  6 hydrogen levels $1s\, , 2s\, , 2p\, , 3s\, , 3p\, , 3d$ with $l_z=0$, modulated by the factors  $\exp(-\beta_i z)$
\begin{multline}\label{eq:4_02}
\psi_1^{tr}=\exp(-\sqrt{\rho^2+(z-h)^2}-\beta_1 z) \ , \\
\psi_2^{tr}=(\sqrt{\rho^2+(z-h)^2}-2)\,\exp(-\frac{1}{2}\sqrt{\rho^2+(z-h)^2}-\beta_2 z) \ ,\\
 \psi_3^{tr}=(z-h)\,\exp(-\frac{1}{2}\sqrt{\rho^2+(z-h)^2}-\beta_3 z) \ ,\\
\psi_4^{tr}=\left(2[\rho^2+(z-h)^2]-18\sqrt{\rho^2+(z-h)^2}+27\right) \times \\ \times \exp(-\frac{1}{3}\sqrt{\rho^2+(z-h)^2}-\beta_4 z) \ , \\
\psi_5^{tr}=(z-h)\,\left(\sqrt{\rho^2+(z-h)^2}-6\right)\ \times \\ \times \exp(-\frac{1}{3}\sqrt{\rho^2+(z-h)^2}-\beta_5 z) \ ,\\
\psi_6^{tr}=\left(\rho^2-2 (z-h)^2 \right)\exp(-\frac{1}{3}\sqrt{\rho^2+(z-h)^2}-\beta_6 z) \ .
\end{multline}
The variational estimate is implemented via the energy functional minimization with respect to variational parameters $\v \b=\{\b_i\}_{i=1}^{i=6}$ and $ \v c=\{c_i\}_{i=1}^{i=6}$. The energy functional takes the form
\begin{equation}\label{eq:4_03}
E^{tr}[\v \b \, , \v c ]=\frac{\< \v c |A\(\v \b\)| \v c \>}{\< \v c |B\(\v \b\)| \v c \>} \ ,
\end{equation}
where   the matrices  $A\(\v \b\)$ and $B\(\v \b\)$ are determined as follows
\begin{equation}\begin{gathered}\label{eq:4_04}
A_{ij}\(\v \b\)= \\ \int\limits_{z \geq 0} dz \,\rho\, d\rho \ \Big[\frac{1}{2}(\vec\nabla\psi_i^{tr}))(\vec\nabla\psi_j^{tr})-
\frac{\psi_i^{tr}\psi_j^{tr}}{\sqrt{\rho^2+(z-h)^2}} \Big] \ + \\ + \ \frac{\lambda}{2}\,\int\limits_{z =0} \rho\, d\rho \ \psi_i^{tr}\psi_j^{tr} \ ,
\end{gathered}\end{equation}
\begin{equation}\label{eq:4_05}
B_{ij}\(\v \b\)=\int\limits_{z\geq 0} dz \,\rho \,d\rho\; \psi_i^{tr}\psi_j^{tr} \ .
\end{equation}
In the next step we diagonalize the matrix $\< \v c |B| \v c \>$ by finding the eigenvalues  $b_i$ and eigenvectors $|\s_i\>$ of the matrix $B$ with subsequent  replacement
\begin{equation}\label{eq:4_07}
| \v c \>=B^{-1/2}|\v \c \> \ ,
\end{equation}
where
\begin{equation}\label{eq:4_08}
B^{-1/2}=\sum\limits_{i}\frac{1}{\sqrt{b_i}}\,|\s_i\>\< \s_i| \ .
\end{equation}
As a result, the minimization of the functional  (\ref{eq:4_03}) reduces to finding  eigenvalues of the matrix $\tilde{A}$
\begin{equation}\label{eq:4_09}
\det \[ \tilde{A}\(\v \b\) - a \(\v \b\) E \]=0 \ ,
\end{equation}
where
\begin{equation}\begin{gathered}\label{eq:4_10}
\tilde{A}\(\v \b \)=B^{-1/2}A\;B^{-1/2}=\sum\limits_{ij}|\s_i\> {\<\s_i| A | \s_j \> \over \sqrt{b_i b_j}}\<\s_j| \ .
\end{gathered}\end{equation}
The minima of found this way eigenvalues $a_i \(\v \b\)$ with respect to variational parameters $\v \b$ present  the variational estimates for the energy levels, corresponding to linear combinations of the form (\ref{eq:4_01}).

\begin{figure}[ht!]
	\includegraphics[width=\columnwidth]{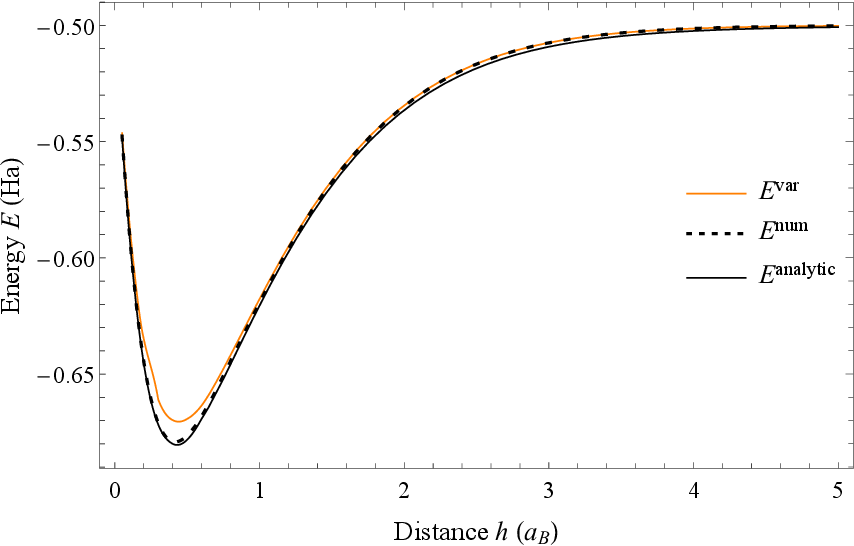}
	\caption{(Color online). Variational estimate based on (\ref{eq:4_09}),(\ref{eq:4_10}), the result of direct numerical calculations in comparison with the analytic answer for the ground state energy, found from (\ref{eq:2_06}), (\ref{eq:2_08}), in dependence on $h$ for the Neumann case with $\l=0$.}
	\label{img:06}
\end{figure}

\begin{figure}[ht!]
	\includegraphics[width=\columnwidth]{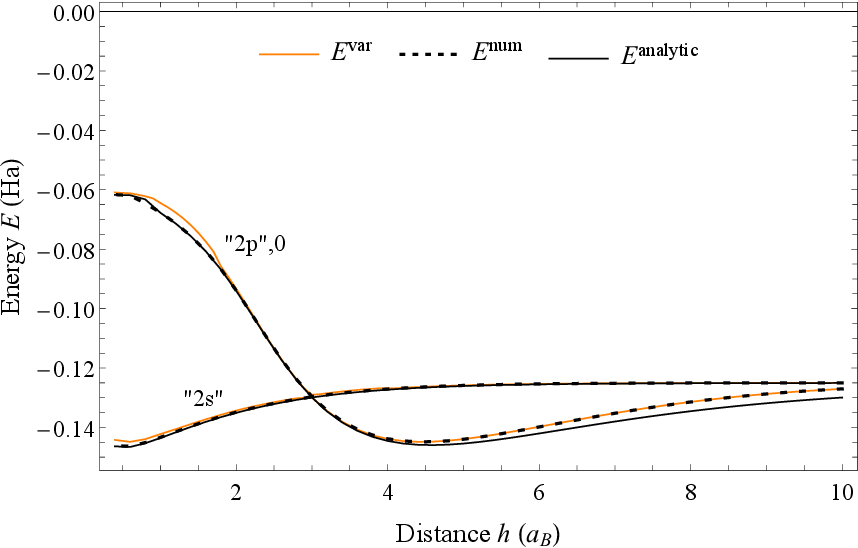}
	\caption{(Color online). Variational estimates based on (\ref{eq:4_09}),(\ref{eq:4_10}), the results of direct numerical calculations in comparison with the analytic answers for two first excited levels with $m=0$, found from (\ref{eq:2_06}), (\ref{eq:2_08}),  in dependence on $h$ for the Neumann case with $\l=0$.}
	\label{img:07}
\end{figure}

\begin{figure}[ht!]
	\includegraphics[width=\columnwidth]{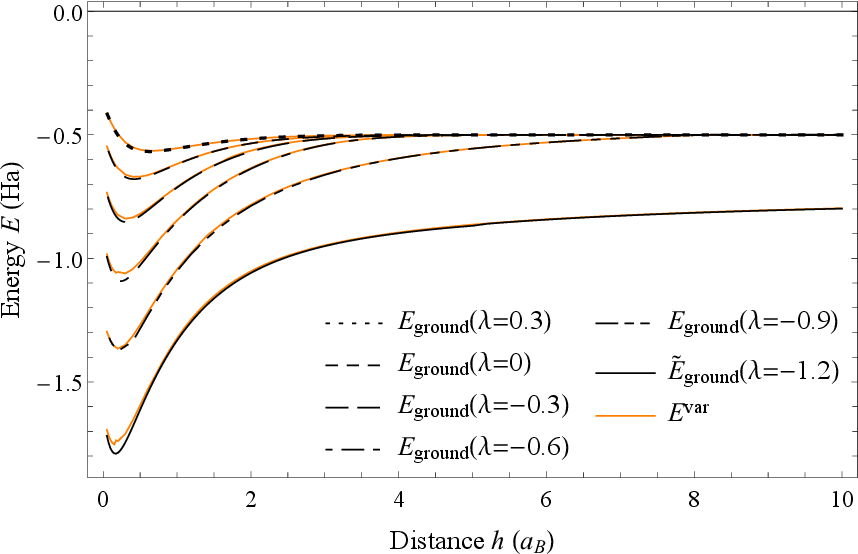}
	\caption{(Color online). $E(h)$ for the ground state and $\lambda=0.3\, , 0\, , -0.3\, , -0.6\, , -0.9\, , -1.2$. Variational estimates vs the results of direct numerical calculations.}
	\label{img:08}
\end{figure}

\begin{figure}[ht!]
\subfigure[]{
		\includegraphics[width=\columnwidth]{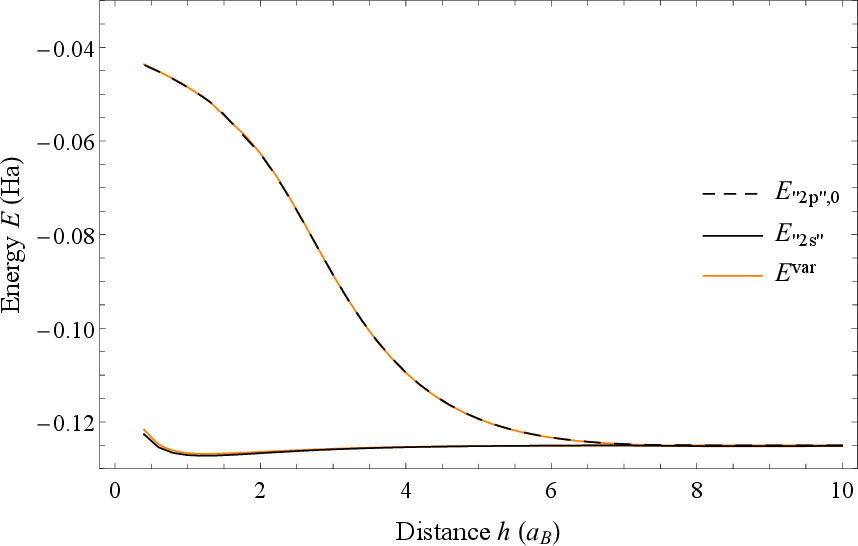}
}
\subfigure[]{
		\includegraphics[width=\columnwidth]{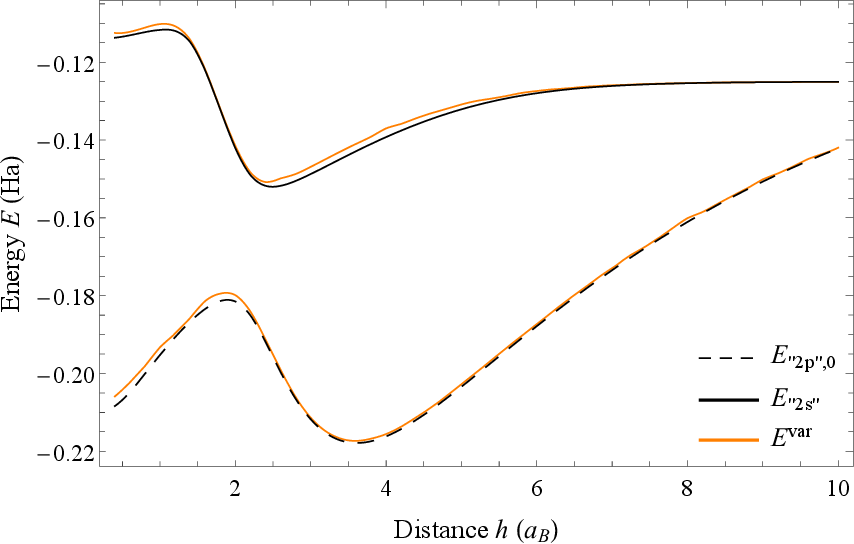}
}
\caption{(Color online). $E(h)$ for the first two excited levels with $m=0$ for (a): $\l= 0.3$; (b): $\l= -0.3$. Variational estimates vs the results of direct numerical calculations. }
\label{img:09}	\end{figure}

The results of calculations based on (\ref{eq:4_09}),(\ref{eq:4_10}) in comparison with the direct numerical analysis and exact answers in corresponding partial cases are presented in Figs.\ref{img:06}-\ref{img:09}. In Figs.\ref{img:06},\ref{img:07} the variational estimates based on (\ref{eq:4_09}),(\ref{eq:4_10}) and the results of direct numerical calculations in comparison with the analytic answers for the ground state energy and two first excited levels, found from (\ref{eq:2_06}), (\ref{eq:2_08}),  are shown in dependence on $h$ for the Neumann case with $\l=0$. From Fig.\ref{img:07} it should be clearly seen that  the trial functions of excited levels with $m=0$, constructed as linear combinations (\ref{eq:4_01}),(\ref{eq:4_02}), are exactly orthogonal to the ground state WF, since in Fig.\ref{img:07} the corresponding variational curves of excited levels lie  always higher than the numerical and exact analytic ones.

In Fig.\ref{img:08} the ground state level is shown for $\lambda=0.3\, , 0\, , -0.3\, , -0.6\, , -0.9\, , -1.2$ in dependence on the distance $h$. There follows from Fig.\ref{img:08} that for $- 0.9 \leq \l \leq + 0.3$ the asymptotics of the ground state for $h \to \inf$ is exponential and tends to the  ``normal'' $1s$-level of free H with $E_{1s}=-0.5\; Ha$, whereas for $\lambda=-1.2$ it represents a power-like one with the limiting energy $-\lambda^2/2=-0.72\, Ha$. It is also clearly seen that in the case of exponential asymptotics  for $- 0.9 \leq \l \leq + 0.3$ the ground state levels approach closely the value  $E_{1s}$ already for  $4 \lesssim h \lesssim 8\;a_B$, whereas for $\l=-1.2$  the power-like ground state level  lies substantially below its asymptotical value even for $h \simeq 10\;a_B$. It should be also mentioned that the minima of all the curves shown in Fig.\ref{img:08} are well-pronounced, but at the same time they lie very close to the border and so are actually indistinguishable against the background of inhomogeneities in atomic layers on the boundary surface (for a more accurate representation of what is meant here see, e.g., Ref.\cite{Bechstedt2012}, Fig.1.1.)

In Figs.\ref{img:09} the dependence $E(h)$ is shown for two first excited levels with $\l=0.3,-0.3$. The case $\lambda=0$ for these levels is already presented in Fig.\ref{img:02}. There follows from these drawings that the asymptotics of excited levels for $h \to \inf$ coincides with those of free H, but if for $\lambda=0.3$ the levels approach their asymptotical values already for $h\simeq 7\;a_B$, in the case of $\lambda=0$ this situation occurs for $h\simeq 10\;a_B$, whereas for $\lambda = -0.3$ it comes out of the considered range of $h$.

It would be also worth to remark that from Figs.\ref{img:06}-\ref{img:09} there follows  that  the variational estimate is worse reproducing the results of numerical calculations, the more negative the value of $\lambda$ and the closer the  nucleus is to the plane. At the same time, the choice of the trial function in the form (\ref{eq:4_01}),(\ref{eq:4_02}) permits to reproduce the effect of hybridization of excited levels for $m=0$ and $\lambda=0$, which has been detected earlier analytically.
Note also that in Figs.\ref{img:06}-\ref{img:09} all the variational curves, corresponding to excited levels, lie above the direct numerical results. It isn't, however,  the general case, since such a choice of trial functions doesn't provide automatically their orthogonality to the ground state WF and so doesn't prevent the situation, when the variational curves lie below the numerical ones due to admixture of the ground state WF.
\begin{figure}[ht!]
\subfigure[]{
		\includegraphics[width=\columnwidth]{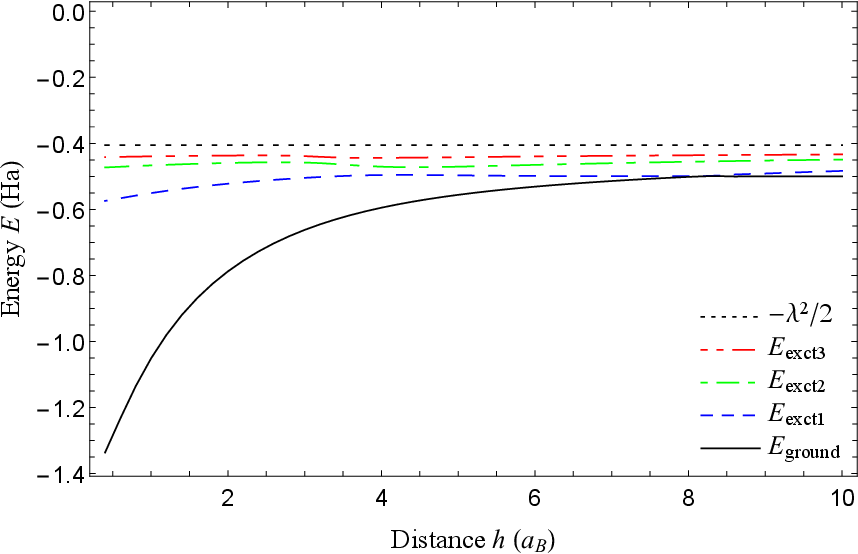}
}
\subfigure[]{
		\includegraphics[width=\columnwidth]{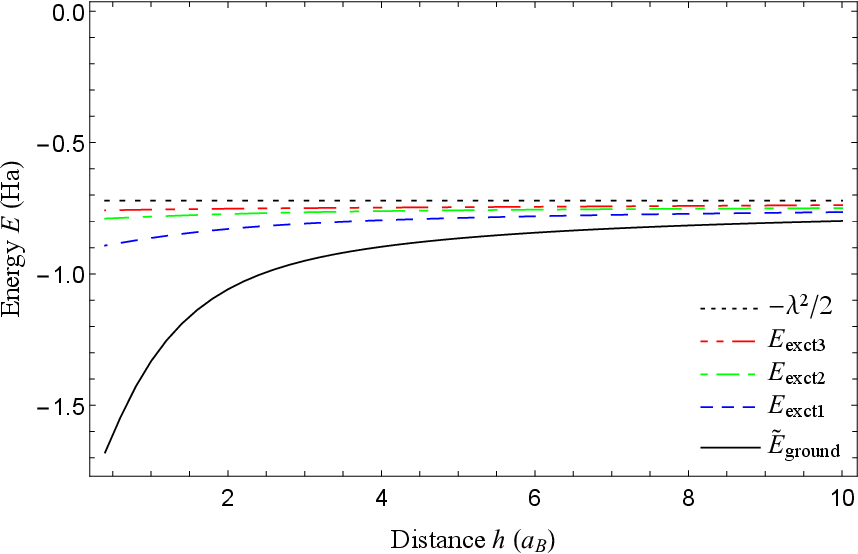}
}
\caption{(Color online). The dependence on $h$ of four first levels with $m=0$ for (a): $\lambda=-0.9$; (b): $\l=-1.2$. Variously dashed colored curves correspond to direct  numerical calculations for excited states, solid line denotes the ground state level, while the dotted line $= -\lambda^2/2$. }
\label{img:10}	\end{figure}

In Figs.\ref{img:10} the dependence on $h$ of  first four levels with $m=0$  is shown for $\lambda=-0.9, -1.2$. For such $\l$ the variational estimates, which are quite effective in description of the ground state level even in the power-like case with $\l=-1.2$ (see Fig.\ref{img:08}), cannot provide the same quality for the excited power-like ones. Therefore in Figs.\ref{img:10} there are presented solely the results, achieved via direct numerical calculations. In Fig.\ref{img:10}a with  $\lambda=-0.9$ the lowest level is the exponential one, which approaches already for $h\simeq 7\;a_B$ its asymptotic value $-0.5\;Ha$, whereas three excited levels are  power-like and  tend very slowly to their common limiting point $-\l^2/2=-0.405\; Ha$. There is no intersection of levels, although for $h \simeq 8\; a_B$ the ground state and the first excited one lie very close to each other.  In Fig.\ref{img:10}b for $\lambda=-1.2$ both the lowest level and the excited ones are  power-like  with the common limiting point $-\l^2/2=-0.72\; Ha$. Note that in the power-like case we avoid to classify the excited levels even in the asymptotics for $h \to \inf$, since in $\Re^3/2$ there remains only $l_z$ as a conserved quantity, and hence, the levels cannot be labeled with  definite values of the orbital moment.  In addition, Figs.\ref{img:10} demonstrate  quite explicitly the general property of the levels with power-like asymptotics that they  approach their limiting point $(-\l^2/2)$ for $h \to \inf$ always much later than the exponential ones.
\begin{figure}[]
	\includegraphics[width=\columnwidth]{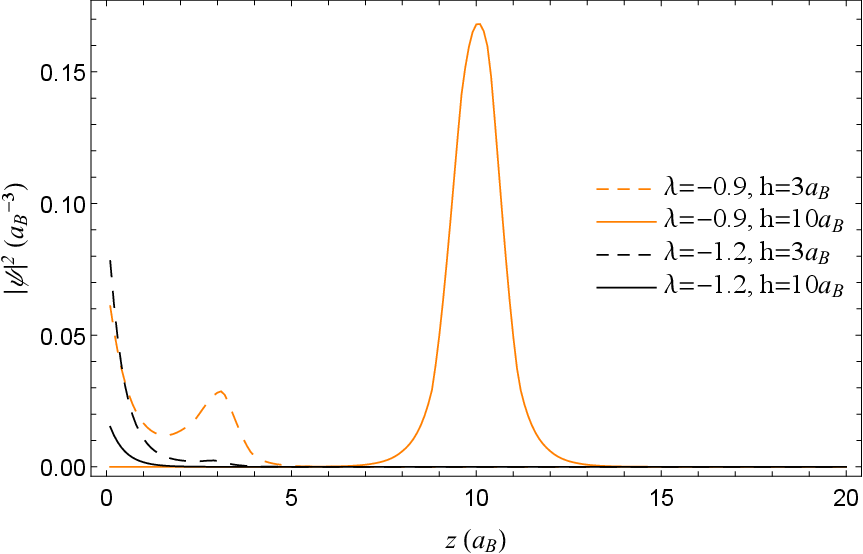}
	\caption{(Color online). The profile of the ground state $|\p|^2$ for $\r=0$.}
	\label{img:11}
\end{figure}
\begin{figure}[]
	\includegraphics[width=\columnwidth]{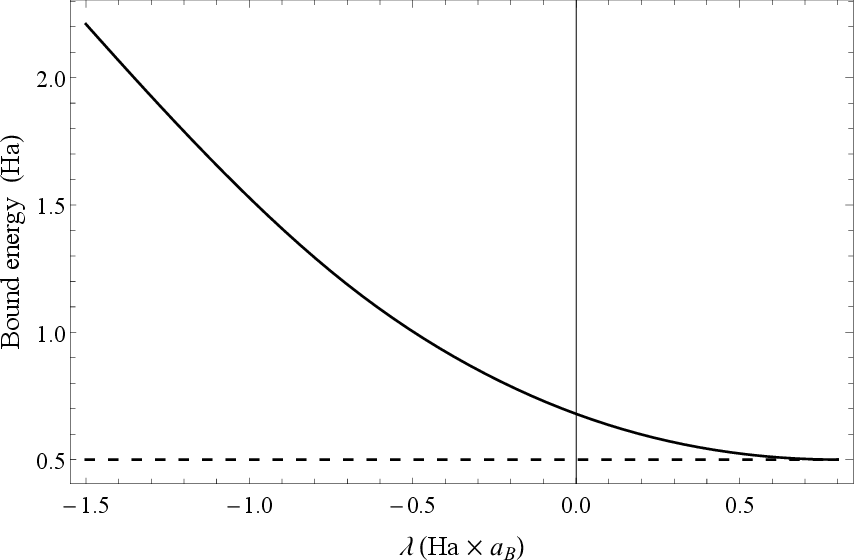}
	\caption{ The dependence of the ground state maximal bound energy on $\l$.}
	\label{img:12}
\end{figure}

In Fig.\ref{img:11}  the profile of the ground state $|\p|^2$, considered  as a function of $z$ for $\r=0$, is shown for two quite representative values of $\l=-0.9\, , -1.2$, and two values of the nucleus position over plane $h=3 a_B\, , 10a_B$. In this case due to the axial symmetry the main contribution to the ground state WF should be formed by the planar $s$-wave component, and the choice $\r=0$ permits to consider indeed this dominating part of WF. From Fig.\ref{img:11} it is clearly seen that for $\l=-0.9$ and $h=10 a_B$, i.e. when the ground state is represented by an exponential level,  $|\p|^2$ is localized in the vicinity of the nucleus and is quite similar to the ``normal'' electronic ground state of free H. For the same $\l$ and $h=3 a_B$ apart from the pick at the nucleus position there appears a tail, smeared over the boundary plane due to sufficiently strong attraction between the electron and the border. Actually, this tail underlies the emergence of power-like ground state levels for  $\l< -1$, whose structure, as it was already mentioned above, contains a large number of spherical harmonics in the form  of a  ``drop'' stuck and partially spread on the boundary plane. Therefore for $\l=-1.2$ the picture is quite different. In this case  the ground state $|\p|^2$ corresponds to a power-like level and is smeared in the vicinity of the border without any pronounced picks at the nucleus position. Another point is that the larger the nucleus position $h$ over the plane, the more pronounced the spreading of $|\p|^2$ over the boundary plane, since in the competition between  electron-nucleus and electron-border interactions, the last one wins.

The dependence of the maximal ground state  bound energy  in the range  $-1.5 < \l < 1$, where the minima of the energy curves are clearly visible (see Fig.\ref{img:08}), is shown in Fig.\ref{img:12}. There follows from the curve, representing this quantity at a given interval, that the general dependence on $\l$ is nonlinear. For more negative $\l$  the maximal bound state energy is reached for sufficiently more small distances $h \ll a_B$, when the electronic WF is localized a  small neighborhood of the boundary plane. In this region the direct numerical calculations become cumbersome, and so the variational estimates via trial function
 come into play. The most pertinent trial function for the ground state of H with  nucleus very close to the border is quite simple, namely
\begin{equation}\label{eq:4_11}
\psi^{tr}_{ground}=N^{-1/2} \exp(-|\l| z - \s r) \ ,
\end{equation}
with $N$ being the normalization coefficient
\begin{equation}\label{eq:4_12}
N= {\pi \over 4 }\, {|\l|+ 2 \s \over \s^2\, (|\l|+ \s)^2}  \ ,
\end{equation}
while $\s > 0$  is the variational parameter.
Such a choice for $\psi^{tr}_{ground}$ is automatically consistent with the boundary condition (\ref{eq:1_03}) and effectively describes  the electronic state  in the form  of a  ``drop'' stuck and partially spread on the border plane, that should be expected in the case of substantially negative $\l$.

\begin{figure}[ht!]
	\includegraphics[width=\columnwidth]{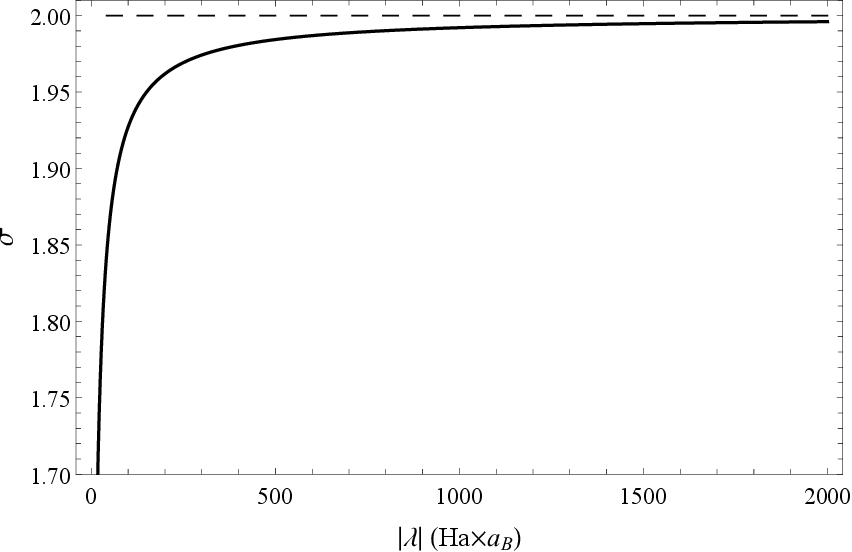}
	\caption{The dependence on $|\l|$ of the parameter $\s$  in $\psi^{tr}_{ground}$, which corresponds to the trial  ground state with the maximal bound energy.}
	\label{img:13}
\end{figure}
Upon substituting (\ref{eq:4_11}) into (\ref{eq:3_01}) with $ h \to 0$ we obtain a very simple expression for the mean value of the ground state energy
\begin{equation}\label{eq:4_13}
E^{tr}_{ground}=-{\l^2 \over 2} + { \s^2 \over 2} - 2 \s { |\l|+\s \over |\l|+ 2 \s } \ .
\end{equation}
Variation of  $E^{tr}_{ground}$ yields the following equation for $\s$
\begin{equation}\label{eq:4_14}
4 \s^3 + 4 \s^2 (|\l|-1) + \s\, |\l| (|\l|-4) - 2 \l^2=0 \ .
\end{equation}
The eq. (\ref{eq:4_14}) possesses one real root, two remaining are complex. So from  (\ref{eq:4_14}) we find for $\s$ a unique answer in the form
\begin{equation}\label{eq:4_15}
\sigma=\frac{t}{6} + \frac{4 |\lambda| ^2+16 |\lambda| +16}{24 t}+\frac{1-|\lambda| }{3} \ ,
\end{equation}
where
\begin{equation}\label{eq:4_16}
\begin{gathered}
t=\left(|\lambda| ^3 + 33 |\lambda| ^2+12 |\lambda| + 8  \ + \right. \\ \left. + \ 3 \sqrt{3} \sqrt{2 |\lambda| ^5+39 |\lambda| ^4+24 |\lambda| ^3+16 |\lambda| ^2}\right)^{1/3} \ .
\end{gathered}
\end{equation}
Moreover, the expression  (\ref{eq:4_15}) defines indeed the minimum of $E^{tr}_{ground}$, since there follows from (\ref{eq:4_13}), that $E^{tr}_{ground}$ as a function of $\s$ behaves like a distorted and shifted parabola. For $|\l| \gg 1$
\begin{equation}\label{eq:4_17}
\sigma \to 2-\frac{8}{|\lambda| }+\frac{80}{|\lambda| ^2}-\frac{960}{|\lambda| ^3}+O\left[\frac{1}{|\lambda| }\right]^{4} \ .
\end{equation}

\begin{figure}[]
	\includegraphics[width=\columnwidth]{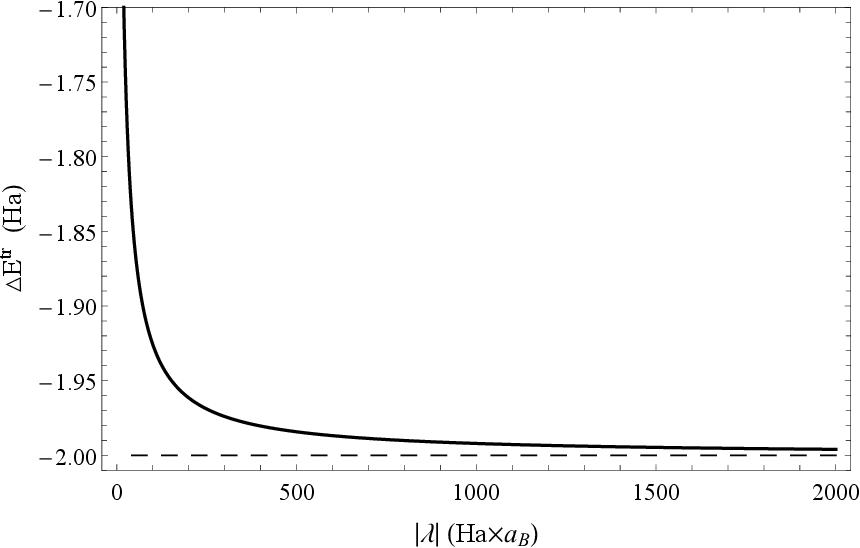}
	\caption{ The dependence on $|\l|$ of the trial ground state  energy vs asymptotics for $h \to \inf$.}
	\label{img:14}
\end{figure}

The results for $\s$ and the minimum of $E^{tr}_{ground}$, found this way, are shown in Figs.\ref{img:13}-\ref{img:14} in dependence on $|\l|$. Remark, that in Fig.\ref{img:14} we present not $E^{tr}_{ground}$ as itself, but the difference between  $E^{tr}_{ground}$ and the asymptotic value $E_{ground}(h \to \inf)=-\l^2 / 2$, namely
\begin{equation}\label{eq:4_18}
\D E^{tr} = E^{tr}_{ground} + \l^2/2 \ .
\end{equation}
This answer is physically more informative, since the main interest is indeed the shift of the ground state level for $h \to 0$ relative to the asymptotics for $h \to \inf$. For $|\l| \gg 1$
\begin{equation}
\D E^{tr} \to -2+\frac{8}{|\lambda| }-\frac{64}{|\lambda| ^2}+\frac{640}{|\lambda| ^3}+O\left[\frac{1}{|\lambda| }\right]^4 \ .
\end{equation}

There follows from Fig.\ref{img:14} that for sufficiently negative $\l$ or, equivalently, large positive electron-plane affinity, and $h \ll a_B$ the ground state level should  lie  more than $ 2\,Ha$  lower than its asymptotics far from the boundary.
The corresponding energy curve, considered  as a function of  $h$,  will be presented by a monotonic power-like one, interpolating  these two asymptotics in a way, quite similar to the ground state curve shown in Fig.\ref{img:10}b.

\section{Conclusion}\label{sect:5}

To summarize, we have shown that for a wide range  of  the ``not going through the border'' parameters the effective atomic potential, treated as a function of the distance $h$ from H to the boundary plane,  reveals in some cases  a well pronounced minimum  at certain finite but non-zero $h$, which corresponds to the mode of ``soaring'' of the atom over the plane.  In other words, we obtain at the atomic level the  microscopic version of the phenomenon called as the ``Mahomet's coffin'' (such phenomenon is well-known in the superconductivity as one of the most prominent manifestations of the Meissner effect) \footnote{To avoid disappointing misunderstandings and speculations the authors would like to underline that the terminology used has long been adopted to denote such a physical effect of soaring over a plane and has nothing to do with theological views.}. However, it would worth to emphasize that due to the boundary  surface roughness such an effect can be an observable one only in those cases when the distance between the energy minimum and the surface is not less than $a_B$. Otherwise, such minima will be indistinguishable against the background of inhomogeneities in atomic layers on the boundary surface.
At the same time, when such a  minimum of atomic level is located at a sufficient distance from the surface (as in Figs.\ref{img:02},\ref{img:04},\ref{img:05},\ref{img:07},\ref{img:09}), so that its roughness becomes insignificant, the soaring mode could take place. The atom in this mode  is able to move freely parallel to the boundary plane  with an arbitrary wavevector ${\v K}_{||}$. If this movement can be stopped, there appears a specific version of a Penning trap for atomic H without complicated 3-dimensional configuration of external fields. The stability of such ``soaring'' states depends on the  overlap of their electronic WF's, since each WF should be more or less located in the vicinity of the nucleus position at the  minimum of the corresponding energy curve, and for well pronounced energy minima could be quite high.

Now let us turn to the general features of the power-like levels, which for sufficiently large positive affinity of the atom to the boundary plane, i.e. for  $\l \leq -1$, turn out to be the lowest ones, and hence, the most important. First it should be mentioned that although in these states the electronic WF is located mostly in the vicinity of the border, these levels are principally different from the other known single-electron surface states. Between the latter in the first place there are  the Tamm states, which arise on the surface of the crystal solids of finite size (see, e.g., Ref.\cite{Cottam2005} and citations therein). But their emergence is intimately connected with the intrinsic structure of the medium, whereas the origin of the power-like levels is the result of specific interplay between the Robin boundary condition on the border and the electron-nucleus electrostatic interaction outside the medium. More closer in nature to the power-like levels there are the ``surface-localized states'', considered in Ref.\cite{Babiker1981}, since their appearance is also caused by the electrostatic interactions in the system. In this case the crucial role is played by the interaction between the atom and medium, filling the half-space $z<0$, when the dielectric properties of the latter are taken into account. Namely, when the dielectric constant $\ve$ of the medium is large, for $h \ll a_B$ the superposition of the atomic potentials and the dielectric response of  medium leads to a potential similar to that of the ``one-dimensional hydrogen atom'' \cite{Loudon1959, Elliott1960} in z-direction
\begin{equation}\label{eq:5_01}
V_S(z)=-\g(\ve)/4z \ ,
\end{equation}
where $\g(\ve)=(\ve-1)/(\ve+1) \simeq 1 $ for $\ve \gg 1$.
However, the interpretation of this result is ambiguous, since the corresponding spectral problem is not self-adjoint and requires additional restrictions to provide a self-adjoint extension, which is in principle not unique \cite{Gitman2012}. In Ref.\cite{Babiker1981} there was proposed the solution by means of the Dirichlet condition on the surface $z=0$ (it is so-called Loudon's extension). In this case the potential (\ref{eq:5_01}) yields the single-electron ``surface-localized states'', which are capable of free motion parallel to the surface, but localized in z-direction in the vicinity of the boundary plane with bound energies
\begin{equation}\label{eq:5_02}
{\cal {E}}_n = -\g(\ve)^2/32\, n^2 \simeq -0.85 \g(\ve)^2/n^2 \  \text{eV} \ , \quad n=1\, , 2\, \dots \ .
\end{equation}
However, such a picture has nothing to do with the power-like levels, since it emerges only for $\l \to \inf$ and  only if the nucleus approaches the border, i.e. when $h \to 0$.  To the contrary, the power-like levels appear in the case of Robin condition with $\l <0$ and exist  for any $0 \leq h \leq \infty$. It should be also mentioned that the combination of the Robin condition with the potential (\ref{eq:5_01}) requires a separate consideration without going to the limit $h \to 0$.

At the same time, indeed these power-like levels define the effective atomic H potential in the case of a large positive affinity. The most important properties  of this potential are the following. First,  it is strictly attractive and long-range, since these levels tend to their limiting values very slow, actually as $\sim 1/h$, in contrast to the van der Waals potential $V_A(h)$ between neutral atoms, which falls down as $\sim 1/h^6$. In Refs.\cite{Sveshnikov2013c,Sveshnikov2013d} it was shown that even for the first critical  $\l_{crit,1}=-1$, when the limiting value of the lowest power-like level coincides with $E_{1s}$, such a  state should be energetically favorable compared to the free atom  up to actual nanocavities with sizes $\sim 100-1000$ nm. Another attractive feature is that although such levels don't provide well-pronounced minima at $h \gtrsim a_B$, for small $h \ll a_B$ their bound energy could exceed several $Ha$  (see Figs.\ref{img:10},\ref{img:12},\ref{img:14}). Therefore, when H in such a power-like  state moves from  the region $h \gg a_B$ to the border, a significant amount of energy can be released,  the more, the closer to the border  the nucleus could be, that can be achieved for sufficiently negative $\l$ or, equivalently, large positive affinity. The magnitude of this energy release, as it follows from Fig.\ref{img:14}, should be estimated as not less than $2 Ha$. Moreover, this effect  can be sufficiently enhanced by the fact that the  attraction of atoms to the plane will be long-range,  and hence, the number of atoms involved in this process can be quite large. It would be worthwhile to note that in a more realistic situation an important role should be played  by the degree of roughness of the boundary  surface, since it determines how close the atom can approach the border and, simultaneously, it influences the magnitude of the affinity and so the value of $\l$. In any case, however, it can be assumed that for a fresh sample with a clean smooth surface and a very low initial H-concentration inside, hence, with a large positive affinity to H, in the first stages of hydrogenation the energy release can be such that it could explain the known events of thirty years ago (the authors believe that their hint is transparent enough to do it without references).

 Finally, it should be mentioned that here we consider a  model stationary picture, which  ends with the adsorption of H on the boundary plane. In a more realistic situation the process continues further  in the form of  physi/chemi-sorption through the boundary surface, for which it is necessary to introduce into $\l$ an imaginary part to provide  the non-vanishing current through the border. In this case we should deal with a  non-stationary picture in terms of inelastic process with metastable states, which imply another techniques including complex energies and Jost functions and therefore will be considered separately.


\vskip 3 true mm

\twocolumngrid

\bibliography{biblio/overplane}

\end{document}